\documentclass[a4paper,11pt]{article}

\usepackage{jheppub}
\usepackage{amssymb}
\usepackage{amsfonts}
\usepackage{amsmath}
\usepackage{amsthm}
\usepackage{mathrsfs}
\usepackage{enumerate}
\usepackage{tikz}
\usepackage[toc, page]{appendix}
\usepackage{comment}
\usepackage{graphicx}
\usepackage{braket}
\usepackage{enumerate}
\usepackage{hhline}
\usepackage{cancel}
\usepackage{enumitem}
\usepackage{xcolor}
\setlist[itemize]{leftmargin=4.5mm}
\usepackage{adjustbox}
\usepackage{multirow}
\usepackage{lscape}
\usepackage{placeins}
\usepackage{tikz}
\usetikzlibrary{arrows,decorations.pathmorphing,backgrounds,fit,positioning,shapes.symbols,chains}
\usepackage{stackengine}
\stackMath
\usepackage{titlesec}
\usepackage{etoolbox}
\makeatletter
\patchcmd{\ttlh@hang}{\parindent\z@}{\parindent\z@\leavevmode}{}{}
\patchcmd{\ttlh@hang}{\noindent}{}{}{}
\makeatother

\titlelabel{\thetitle.\!\!\quad}
\let \savenumberline \numberline
\def \numberline#1{\savenumberline{#1.}}
\makeatletter
\def\@fpheader{\relax}
\makeatother

\newcommand\tsup[2][2]{%
 \def\useanchorwidth{T}%
  \ifnum#1>1%
    \stackon[-.5pt]{\tsup[\numexpr#1-1\relax]{#2}}{\scriptscriptstyle\sim}%
  \else%
    \stackon[.5pt]{#2}{\scriptscriptstyle\sim}%
  \fi%
vv}

\newcommand{\tl}{\tau}
\newcommand{\tlV}{\tau}
\newcommand{\ttr}{H}
\newcommand{\ttrV}{E}
\newcommand{\tB}{B}

\newcommand{\p}{\partial}

\newcommand{\CM}{\mathcal{M}}

\newcommand{\Ap}{{A'}}
\newcommand{\Bp}{{B'}}
\newcommand{\Cp}{{C'}}
\newcommand{\qd}{{\mathcal{D}}}

\newcommand{\lr}{\left(}
\newcommand{\rr}{\right)}
\newcommand{\ls}{\left[}
\newcommand{\rs}{\right]}
\newcommand{\be}{\begin{equation}}
\newcommand{\ee}{\end{equation}}

\allowdisplaybreaks[1]

\title{\ \vspace{1.6cm} \\
String Theory and String Newton-Cartan Geometry}
\author{Eric A.~Bergshoeff${}^{\, a}$, Jaume Gomis${}^{\, b}$, Jan Rosseel${}^{\, c}$, Ceyda \c{S}im\c{s}ek${}^{\, a}$, Ziqi Yan${}^{\, b}$}
\emailAdd{E.A.Bergshoeff@rug.nl}
\emailAdd{jgomis@perimeterinstitute.ca}
\emailAdd{\,\,\,\,\,\,\,\,\,\,\,\,\,\,\,\,\,\,\,\,\,\,jan.rosseel@univie.ac.at}
\emailAdd{c.simsek@rug.nl}
\emailAdd{\,\,\,\,\,\,\,\,\,\,\,\,\,\,\,\,\,\,\,\,\,\,zyan@perimeterinstitute.ca}
\affiliation{
${}^a$Van Swinderen Institute, University of Groningen,\\ Nijenborgh 4, 9747 AG Groningen, The Netherlands\medskip\\
${}^b$Perimeter Institute for Theoretical Physics,\\ 31 Caroline St N, Waterloo, ON N2L 6B9, Canada\medskip\\
${}^{c}$Faculty of Physics, University of Vienna,\\ Boltzmanngasse 5, A-1090, Vienna, Austria\\[.5truecm]}
\abstract{Nonrelativistic string theory is described by a sigma model with a relativistic worldsheet and a nonrelativistic target spacetime geometry, that is called string Newton-Cartan geometry.
In this paper we obtain string Newton-Cartan geometry   as a limit of the Riemannian geometry of General Relativity  with  a fluxless two-form field.
We then apply the same limit to relativistic string theory  in curved background fields and show that it leads to nonrelativistic string theory in a string Newton-Cartan geometry coupled to a Kalb-Ramond and dilaton field background. Finally, we use our limiting procedure to study the spacetime equations of motion and the T-duality transformations of nonrelativistic string theory. Our results reproduce  the recent studies of beta-functions and T-duality of nonrelativistic string theory obtained from the microscopic  worldsheet definition of nonrelativistic string theory.\\
\vskip 1truecm

\centerline{\emph{Dedicated to the memory of P.G.O. Freund.}}}

\begin{document}

\maketitle

\section{Introduction}

Formulating a self-consistent theory of quantum gravity is perhaps one of the most outstanding problems in modern theoretical physics. String theory provides a quantum framework that successfully incorporates massless spin-two degrees of freedom and in which fundamental questions on quantum gravity can be addressed. While string theory is reasonably well understood as a perturbative theory, a proper comprehension of its non-perturbative aspects is however still lacking. In this regard, it is useful to study limits in which string theory is considerably simplified, as such studies can give further hints about what the full non-perturbative formulation of string theory looks like.

One such limit was introduced in \cite{Seiberg:2000ms, Gopakumar:2000na, Klebanov:2000pp}, which was then formulated by Gomis and Ooguri in \cite{Gomis:2000bd} as a two-dimensional quantum field theory that is consistent, unitary and ultra-violet complete and that was dubbed nonrelativistic string theory.\,\footnote{See also \cite{Danielsson:2000gi}.} This theory contains extra worldsheet one-forms beyond the usual worldsheet fields that correspond to coordinates on the target spacetime. The inclusion of these one-forms in \cite{Gomis:2000bd} endows the target spacetime of dimension $d$ with a foliation structure, that splits the spacetime into a two-dimensional longitudinal and a $(d-2)$-dimensional transverse sector. Such a spacetime is characterized by a type of boost symmetries, called string-Galilean boosts, under which the $d-2$ transverse directions are transformed into the two longitudinal directions, but not vice versa. The associated string spectrum is nonrelativistic and invariant under these string-Galilean boosts. The spacetime geometry of nonrelativistic string theory was  studied in a series of papers \cite{Gomis:2005pg, nrGalilei, NCbranes, stringyNC} and was argued more recently in \cite{Bergshoeff:2018yvt} to correspond to the so-called string Newton-Cartan geometry.\footnote{For other recent work on nonrelativistic strings, see \cite{Batlle:2016iel, Gomis:2016zur, Batlle:2017cfa, HHO, Kluson, Kluson:2018grx, Harmark:2018cdl, Kluson:2018vfd, Kluson:2019ifd, Roychowdhury:2019vzh, Roychowdhury:2019qmp}. In \cite{HHO, Kluson, Harmark:2018cdl}, for zero torsion, a specific truncation of string Newton-Cartan gravity (with zero $B$-field and dilaton) in the target space was considered, which leads to Newton-Cartan gravity in one dimension lower, supplemented with an extra worldsheet scalar parametrizing the spatial foliation direction. A more thorough examination of this relation has been put forward in \cite{Harmark:2019upf}.\label{footnote:nrst}}

In this paper, we start with a detailed study of how string Newton-Cartan geometry appears as a nonrelativistic limit of the Riemannian geometry of General Relativity (GR) with a fluxless two-form field. Two features of this limit are worth pointing out. First, there is an ambiguity that appears when taking this limit: different limits parametrized by arbitrary functions lead to the same string Newton-Cartan geometry. This ambiguity will play an important role in applications to nonrelativistic string theory later in this paper.
Secondly, while it is well-known that in GR all components of the spin connection can be determined in terms of Vielbeine as solutions of zero torsion constraints, this is no longer true after taking the nonrelativistic limit. Indeed, we will see that the limit of the zero torsion constraints of GR leads to zero torsion constraints for string Newton-Cartan geometry that leave some components of the spin connections undetermined.\footnote{The same phenomenon has appeared in \cite{Bergshoeff:2018vfn}, where an action was constructed for a four-dimensional string Newton-Cartan gravity theory, that gives dynamics to an extended version of string Newton-Cartan geometry. In this theory, there are also undetermined spin connection components in the second order formalism, and these play the role of Lagrange multipliers that impose some of the components of the foliation constraints of string Newton-Cartan geometry. (Also see \cite{Aviles:2019xed} for a closely related case in which an action principle for stringy Galilei Newton-Hooke Chern-Simons gravity in $(2+1)$-dimensions is put forward.) In general, this type of unconstrained spin connections usually appears as auxiliary fields in the spacetime formalism and does not show up in the worldsheet formalism.}

With the sharpened understanding of string Newton-Cartan gravity, we then consider nonrelativistic string theory as a zero slope, near critical electric field limit of relativistic string theory in curved backgrounds with a Kalb-Ramond and dilaton field.\footnote{See also \cite{Gomis:2005pg}.} This limit is closely related to the noncommutative open string (NCOS) limit \cite{Klebanov:2000pp}. Intriguingly, unlike the situation in relativistic string theory, one finds that the background fields that appear in the nonrelativistic string action are only defined up to St\"{u}ckelberg-type of symmetries, when including a Kalb-Ramond and dilaton field \cite{Bergshoeff:2018yvt}. These St\"{u}ckelberg symmetries are related to the fact that different limits parametrized by arbitrary functions lead to the same string Newton-Cartan gravity, as mentioned previously.

We then consider two applications of the techniques developed in this paper: one to spacetime equations of motion and one to T-duality transformations in nonrelativistic string theory. First, 
we apply the limiting procedure to the one-loop beta-functions in relativistic string theory. Setting the resulting beta-functions to zero, we derive the equations of motion that dictate the backgrounds on which nonrelativistic string theory can be consistently defined quantum mechanically. These equations of motion agree with the ones found in \cite{Weyl}, where the quantum Weyl invariance of vertex operators in nonrelativistic string theory has been studied and where the same set of equations of motion has been derived in an intrinsic manner.\,\footnote{For a study of Weyl invariance of the sigma model in a torsional Newton-Cartan background, see \cite{Gallegos:2019icg}.}
Secondly, in \cite{Bergshoeff:2018yvt}, it was proven that T-duality along a longitudinal spatial isometry in string Newton-Cartan geometry gives relativistic string theory on a Lorentzian background with a compact lightlike isometry. This relation establishes a first principles definition of string theory in the discrete light cone quantization (DLCQ), a prevailing approach to nonperturbative string/M-theory that provided the conjecture stating the equivalence of M-theory in DLCQ and Matrix theory \cite{Banks:1996vh, Susskind:1997cw, Seiberg:1997ad}. In this paper, we will show that the same T-duality transformations also arise as a limit of the Buscher rules \cite{Buscher:1987sk, Buscher:1987qj} in relativistic string theory.

This paper is organized as follows. First, in \S\ref{sec:NRstring}, we briefly present the original nonrelativistic string theory in flat spacetime with a special emphasis on the different symmetry algebras that will occur in the rest of this paper. Next, in \S\ref{sec:sncg}, we focus on obtaining string Newton-Cartan geometry and its underlying local spacetime symmetries via a nonrelativistic limit of the Riemannian geometry of GR with a fluxless two-form field.
In \S\ref{sec:appl}, we study nonrelativistic string theory as a limit of relativistic string theory and we discuss the coupling of string Newton-Cartan geometry to nonrelativistic string theory. We will then focus on two special topics: the spacetime equations of motion and nonrelativistic T-duality in nonrelativistic string theory. Finally, we give our conclusions in \S\ref{sec:conclusions}. In the appendices, we present useful results regarding the algebra of global spacetime symmetries of nonrelativistic string theory in flat spacetime. We also discuss the gauging of the subalgebra of these flat spacetime symmetries that remains as a symmetry algebra of the nonrelativistic string action, as an alternative way of constructing string Newton-Cartan geometry.

\section{Nonrelativistic String Theory in Flat Spacetime} \label{sec:NRstring}

One purpose of this paper is to give an overview of string Newton-Cartan geometry and its role in nonrelativistic string theory thereby reviewing some old results as well as providing several new insights. In this introductory section, we give a brief review of nonrelativistic string theory in flat spacetime and its global target spacetime symmetries.

Nonrelativistic string theory is described by a two-dimensional sigma model defined on a Riemann surface $\Sigma$\,, parametrized by $\sigma^\alpha$\,, $\alpha = 0, 1$\,. For now, we apply the conformal gauge and thus take the Riemann surface to be flat with metric $\eta_{\alpha\beta} = \mathrm{diag}(-1,1)$.
The worldsheet fields of nonrelativistic string theory consist of worldsheet scalars parametrizing the spacetime coordinates $x^\mu$\,, $\mu = 0, 1, \cdots, d-1$ and two one-form fields on the worldsheet, which we denote by $\lambda$ and $\overline{\lambda}$\,. We take the decomposition $x^\mu = (x^A, x^\Ap)$\,, with $A = 0,1$ and $\Ap = 2, \cdots, d-1$\,. The nonrelativistic string theory action in flat spacetime is then given by the following sigma model action \cite{Gomis:2000bd,Bergshoeff:2018yvt}
\be \label{eq:actionflat}
	S_\text{flat} = - \frac{1}{4\pi\alpha'} \int d^2 \sigma \bigl( \, \p x^\Ap \overline{\p} x^\Bp \delta_{\Ap\Bp} + \lambda \, \overline{\p} X + \overline{\lambda} \, \p \overline{X} \, \bigr)\,,
\ee
where $\alpha'$ is the Regge slope, that determines the string tension $T$ via
\be
	T = \frac{1}{2\pi\alpha'}\,.
\ee
In \eqref{eq:actionflat}, we introduced spacetime light-cone coordinates
\be
	X = x^0 + x^1\,,
		\qquad
	\overline{X} = x^0 - x^1\,,
\ee
as well as light-cone coordinates on the worldsheet such that
\be \label{eq:dbard}
	\p = \frac{\p}{\p \sigma^0} + \frac{\p}{\p \sigma^1}\,,
		\qquad
	\overline{\p} = - \frac{\p}{\p \sigma^0} + \frac{\p}{\p \sigma^1}\,.
\ee
The one-form fields $\lambda$ and $\overline{\lambda}$ appear in $S_\text{flat}$ as Lagrange multipliers, that impose that $X$ (resp. $\overline{X}$) is holomorphic (resp. anti-holomorphic):
\be \label{eq:holoconds}
	\overline{\p} X = \p \overline{X} = 0\,.
\ee
Similarly, we also have that $\lambda$ (resp. $\overline{\lambda}$) are holomorphic (resp. anti-holomorphic) as a consequence of the equations of motion of $X$ and $\overline{X}$:
\be \label{eq:holocondslambda}
	\overline{\p} \lambda = \p \overline{\lambda}  = 0\,.
\ee

Following \cite{Batlle:2016iel}, it turns out that the action \eqref{eq:actionflat} is invariant under an infinite number of spacetime symmetries, with (anti-)holomorphic parameters $f(X),  \overline{f}(\overline{X}), g_{A'}(X), \overline{g}_{A'}(\overline{X})$ and constant parameters $\Lambda^{A'}{}_{B'}$, given by\footnote{We would like to thank Joaquim Gomis for bringing our attention to the extended Galilean symmetries in \cite{Batlle:2016iel}.}
\begin{subequations} \label{eq:globaltrnsf}
\begin{align}
	\delta x^{A'} & = g^{A'} (X) + \overline{g}^{A'} (\overline{X}) - \Lambda^{A'}{}_{B'} x^{B'} \,,  \\[2pt]
	\delta X & = f (X)\,,
		&
	& \hspace{-5cm} \delta \lambda = - \lambda \, f' (X) - 2 \, g'_{A'} \! (X) \, \p x^{A'}\,, \\[2pt]
	\delta \overline{X} & = \overline{f} (\overline{X})\,,
		&
	& \hspace{-5cm} \delta \overline{\lambda} = - \overline{\lambda} \, \overline{f}' (\overline{X}) - 2 \, \overline{g}'_{A'} \! (\overline{X}) \, \overline{\p} x^{A'}\,,
\end{align}
\end{subequations}
where we have defined
\begin{eqnarray}
f'(X) = \frac{\p f}{\p X}\,, \hskip .5truecm \overline{f}'(\overline{X}) = \frac{\p \overline{f}}{\p \overline{X}}\,,\hskip .5truecm
g'_{A'}(X) = \frac{\p g_{A'}}{\p X}\,,\hskip .5truecm \overline{g}'_{A'}(\overline{X}) = \frac{\p \overline{g}_{A'}}{\p \overline{X}}\,.
\end{eqnarray}
The algebra that these symmetries satisfy is derived in Appendix \ref{app:globalsymm}. 
Later in this paper, we will use a limiting procedure to construct the Polyakov action for nonrelativistic string theory coupled to background fields corresponding to a curved string Newton-Cartan geometry and a Kalb-Ramond and dilaton field. A finite subset of symmetries of this infinite-dimensional algebra will then be realized as sigma model symmetries acting on the background fields.
The algebra of this finite subset will be referred to as the \emph{string Newton-Cartan algebra}; it is an extension of the algebra introduced in \cite{nrGalilei, stringyNC}, that we now refer to as the \emph{string Bargmann algebra}. Historically, the string Bargmann algebra was first introduced as the symmetry algebra underlying nonrelativistic string theory. However, it turns out that the correct symmetry algebra is the string Newton-Cartan algebra \cite{Harmark:2018cdl}. Details on the embedding of these two algebras in the infinite-dimensional algebra can be found in Appendix \ref{app:globalsymm}, while appendices \ref{app:gauging} and \ref{app:gaugingext} are concerned with their gauging as an alternative way of constructing string Newton-Cartan geometry.

\section{String Newton-Cartan Geometry as a Limit of General Relativity} \label{sec:sncg}

In this section, we show how string Newton-Cartan geometry arises when considering a special nonrelativistic limit of General Relativity in the first-order formalism and use this to discuss various aspects of string Newton-Cartan geometry and its underlying local symmetries. Its coupling to the nonrelativistic string and other string-theory related aspects will be discussed in the next section.

\subsection{Elements of General Relativity}

We will start from GR in the Vierbein description, i.e. we will consider a Vierbein field $\hat{E}_\mu{}^{\hat{A}}$\,, with $\hat{A} = 0\,, 1\,, \cdots, d-1$\,, whose inverse is denoted by $\hat{E}^\mu{}_{\hat{A}}$:
\be \label{eq:invE}
	\hat{E}_\mu{}^{\hat{A}} \hat{E}^\mu{}_{\hat{B}} = \delta^{\hat{A}}_{\hat{B}}\,,
		\qquad
	\hat{E}_\mu{}^{\hat{A}} \hat{E}^\nu{}_{\hat{A}} = \delta_\mu^\nu\,.
\ee
In order to define a limit that leads to string Newton-Cartan geometry, we will add a two-form gauge field $\hat{M}_{\mu\nu}$ to the field content. This two-form gauge field is auxiliary in the sense that we constrain it to have zero curvature
\begin{equation}\label{constraint}
	\partial_{[\mu} \hat{M}_{\nu\rho]}=0\,,
\end{equation}
so that it does not describe any propagating degrees of freedom. The above fields transform under
Lorentz transformations, with parameters $\hat{\Lambda}_{\hat{A}\hat{B}}= - \hat{\Lambda}_{\hat{B}\hat{A}}$\,, and gauge transformations, with parameters $\hat{\eta}_\mu$\,, as follows:
\begin{align}\label{Rtransf}
	\delta \hat{E}_\mu{}^{\hat{A}} & =
		\hat{\Lambda}^{\hat{A}}{}_{\hat{B}} \, \hat{E}_\mu{}^{\hat{B}}\,,
		\qquad%
	\delta \hat{M}_{\mu\nu} =
		\partial_{\mu} \hat{\eta}_{\nu} - \partial_{\nu} \hat{\eta}_{\mu} \,.
\end{align}
We furthermore declare that all gauge fields transform as covariant vectors under diffeomorphisms with parameters $\hat{\xi}^\mu$.\footnote{Alternatively, the diffeomorphisms can be introduced in a gauging procedure of the Poincar\'e algebra. In that case, the Vierbein is considered as a gauge field for translations parametrized by $\hat{\Xi}^{\hat{A}}$\,. Once the torsionlessness constraint \eqref{eq:torsionless} is imposed, diffeomorphisms with parameters $\hat{\xi}^\mu$ can be expressed as a combination of translations with parameters $\hat{\Xi}^{\hat{A}} \equiv \hat{\xi}^\mu \hat{E}_\mu{}^{\hat{A}}$ and other symmetries of the theory.}

The Vierbein formulation of GR also includes the spin connection $\hat{\Omega}_\mu{}^{\hat{A}\hat{B}}$, that is a gauge connection for local Lorentz transformations and thus transforms as follows
\begin{equation}
\delta \hat{\Omega}_\mu{}^{\hat{A}\hat{B}} = \partial_\mu \hat{\Lambda}^{\hat{A}\hat{B}} - 2 \hat{\Omega}_\mu{}^{\hat{C}[\hat{A}} \, \hat{\Lambda}^{\hat{B}]}{}_{\hat{C}}\,,
\end{equation}
under local Lorentz transformations. In the second order formulation, the spin connection is not an independent gauge field. Instead, it depends on the Vierbein in such a way that the zero torsion constraint
\be \label{eq:torsionless}
	\p_{[\mu} \hat{E}_{\nu]}{}^{\hat{A}} - \hat{\Omega}_{[\mu}{}^{\hat{A}\hat{B}} \hat{E}_{\nu]\hat{B}} = 0\,,
\ee
is identically satisfied. Solving \eqref{eq:torsionless} for $\hat{\Omega}_\mu{}^{\hat{A}\hat{B}}$ leads to the following expression\,\footnote{Note that later, we will take the limit in the curvature constraint \eqref{eq:torsionless} with an independent spin-connection field and not in the solution \eqref{eq:hatOmege} where the spin-connetion is dependent. Equivalently, one may also take the same limit directly in the second-order formalism.}
\begin{eqnarray} \label{eq:hatOmege}
\hat{\Omega}_\mu{}^{\hat{A}\hat{B}} = -2 \hat{E}_\mu{}^{[\hat{A}\hat{B}]} + \hat{E}_{\mu \hat{C}} \hat{E}^{\hat{A}\hat{B}\hat{C}} \,.
\end{eqnarray}
Here, we defined
\begin{equation}
	\hat{E}_{\mu \nu}{}^{\hat{A}} \equiv \partial_{[\mu} \hat{E}_{\nu]}{}^{\hat{A}}\,,
\end{equation}
and we turned curved $\mu$-indices into flat $\hat{A}$-indices by contracting them with (inverse) Vielbeine, as in e.g.
\be
	\hat{E}_{\mu\hat{A}}{}^{\hat{B}} \equiv \hat{E}^\nu{}_{\hat{A}} \hat{E}_{\mu\nu}{}^{\hat{B}}\,.
\ee
The curvature two-form of $\hat{\Omega}_\mu{}^{\hat{A}\hat{B}}$ is defined as
\be \label{eq:curvature}
	\hat{R}_{\mu\nu}{}^{\hat{A}\hat{B}}(\hat\Omega) = 2 \bigl( \p_{[\mu} \hat{\Omega}_{\nu]}{}^{\hat{A}\hat{B}} + \hat{\Omega}_{[\mu}{}^{\hat{A}\hat{C}} \, \hat{\Omega}_{\nu]}{}^{\hat{B}}{}_{\hat{C}} \bigr)\,.
\ee
The usual Levi-Civita connection $\hat{\Gamma}^\rho{}_{\mu\nu}$ and associated Riemann curvature tensor $\hat{R}^\rho{}_{\sigma\mu\nu} (\hat{\Gamma})$ can then be expressed in terms of the Vierbein and spin-connection as
\be \label{eq:connection}
	\hat{\Gamma}^\rho{}_{\mu\nu} = \hat{E}^\rho{}_{\hat{A}} \bigl[ \p_{(\mu} \hat{E}_{\nu)}{}^{\hat{A}} - \hat{\Omega}_{(\mu}{}^{\hat{A}\hat{B}} \hat{E}_{\nu)\hat{B}} \bigr]\,,
\ee
and
\be \label{eq:Riemann}
	\hat{R}^\rho{}_{\sigma\mu\nu} (\hat{\Gamma}) = - \hat{E}^\rho{}_{\hat{A}} \hat{E}_\sigma{}^{\hat{B}} \hat{R}_{\mu\nu}{}^{\hat{A}}{}_{\hat{B}} (\hat{\Omega})\,.
\ee

\subsection{String Newton-Cartan Geometry from General Relativity} \label{sec:limit}

In this subsection, we will consider a nonrelativistic limit of the kinematical structure and transformation rules \eqref{Rtransf} of GR that will lead to string Newton-Cartan geometry. In order to do this, we expand the relativistic fields $\hat{E}_\mu{}^{\hat{A}}$ and $\hat{M}_{\mu\nu}$ in terms of the fields that will soon define string Newton-Cartan geometry\,\footnote{Note that only after taking the limit $c \rightarrow \infty$\,, the fields at the r.h.s.~of \eqref{eq:Vielbeinexp} may be identified with the fields of string Newton-Cartan geometry.}. Introducing the speed of light $c$, these expressions are given by\,\footnote{We use the convention that $\epsilon_{01} = 1.$}
\begin{subequations} \label{eq:Vielbeinexp}
\begin{align}
&\hat{E}_\mu{}^A = X_\mu{}^A + \frac{1}{c} \, m_\mu{}^A\,, \hskip 1truecm
 \hat{E}_\mu{}^{A^\prime} = E_\mu{}^{A^\prime}\,,\label{RNC1}\\[2pt]
&\hat{M}_{\mu\nu} = - {X}_\mu{}^A {X}_\nu{}^B\, \epsilon_{AB}\hskip 1truecm \textrm{with}\hskip 1truecm {X}_\mu{}^A =
c \, \tau_\mu{}^A - \frac{1}{c} \, C_\mu{}^A \,,\label{RNC2}
\end{align}
\end{subequations}
where we have split the index $\hat{A}$ in a longitudinal index $A=0,1$ and a transversal one $A^\prime = 2,\cdots, d-1$.
In \eqref{eq:Vielbeinexp}, $\tau_\mu{}^A$, $E_\mu{}^{A^\prime}$ and $m_\mu{}^A$ correspond to the independent fields that will define string Newton-Cartan geometry; in particular, $\tau_\mu{}^A$ and $E_\mu{}^{A'}$ will play the role of the longitudinal and transverse Vielbein fields, respectively. The $C_\mu{}^A$ are arbitrary functions, that will not show up in the geometrical objects of the string Newton-Cartan geometry, that will originate from the limiting procedure. They will however play an important role in the formulation of nonrelativistic string theory later in this paper.\footnote{More precisely, the presence of the arbitrary function is related to the fact that the nonrelativistic background fields are determined by equivalence classes only.  In \cite{Kluson:2018uss, Kluson:2019ifd}, the expansion of \eqref{eq:Vielbeinexp} and the $c \rightarrow \infty$ limit in nonrelativistic string theory was studied, but only for a fixed element of the equivalence class, i.e.~for a  fixed expression for $C_\mu{}^A$ given by $C_\mu{}^A = \frac{1}{2} m_\mu{}^A$\,. See also \cite{Bergshoeff:2015uaa} for the Newton-Cartan case.} Later in this section, we  will show that not all  components of the spin connections are dependent.  However, the  independent components  do not appear in the nonrelativistic {string action and drop out in the expressions for the $\beta$-functions. 

Applying the expansions in \eqref{eq:Vielbeinexp} to \eqref{eq:invE}, and then taking the $c \rightarrow \infty$ limit, we find the following projective invertibility conditions on $\tau_\mu{}^A$ and $E_\mu{}^{A'}$\,,
\begin{subequations}
\begin{align}
	\tau^\mu{}_A \tau_\mu{}^B & = \delta^B_A\,,
		&
	\hspace{-2cm}\tau_\mu{}^A \tau^\nu{}_A + E_\mu{}^{\Ap} E^\nu{}_{\Ap} & = \delta_\mu^\nu\,, \\[2pt]
	E_\mu{}^{\Ap} E^\mu{}_{\Bp} & = \delta^{\Ap}_{\Bp}\,,
		&
	\hspace{-2cm}\tau^\mu{}_A E_\mu{}^{\Ap} = E^\mu{}_{\Ap} \tau_\mu{}^A & = 0\,.
\end{align}
\end{subequations}
We also adopt the following expansions of the independent spin connection components,
\begin{subequations} \label{eq:Omegaexp}
\begin{align}
	\hat{\Omega}_\mu{}^{AB} & = \Bigl( \Omega_\mu + \frac{1}{c^2} \, n_\mu \Bigr) \epsilon^{AB} + O(c^{-4})\,, \\[2pt]
	\hat{\Omega}_\mu{}^{AA'} & =  \frac{1}{c} \, \Omega_\mu{}^{AA'} + O(c^{-3})\,, \\[2pt]
	\hat{\Omega}_\mu{}^{A'B'} & = \Omega_\mu{}^{A'B'} + O(c^{-2})\,,
\end{align}
\end{subequations}
as well as similar expansions of the curvature two-form components,
\begin{subequations} \label{eq:Rexp}
\begin{align}
	\hat{R}_{\mu\nu}{}^{AB} & = R_{\mu\nu} (M) \, \epsilon^{AB}+ O(c^{-2})\,, \\[2pt]
	\hat{R}_{\mu\nu}{}^{AA'} & = \frac{1}{c} R_{\mu\nu}{}^{AA'} (G) + O(c^{-3})\,, \\[2pt]
	\hat{R}_{\mu\nu}{}^{A'B'} & = R_{\mu\nu}{}^{A'B'}(J) + O(c^{-2})\,.
\end{align}
\end{subequations}
Here, we introduced a gauge field $n_\mu$ to parametrize the $O(c^{-2})$ term in the expansion of $\hat{\Omega}_\mu{}^{AB}$ in \eqref{eq:Omegaexp}. After taking the $c \rightarrow \infty$ limit, $n_\mu$ will drop out in the resulting string Newton-Cartan geometry. The curvature two-forms $R_{\mu\nu} (M)$\,, $R_{\mu\nu}{}^{AA'} (G)$ and $R_{\mu\nu}{}^{A'B'} (J)$ will turn out to be associated with the longitudinal Lorentz rotation generator $M$, the boost generator $G_{AA'}$ and the transverse rotation generator $J_{A'B'}$, respectively. Their expressions in terms of the spin connections will be derived later in \eqref{eq:curvaturetwoformsMJG}.

In order to obtain a well-defined limit, we have not made the most general expansion of the $\hat{M}_{\mu\nu}$ field up to some order of $c$ but instead we have taken a very specific one.
The above expansions imply the following ones for the inverse Vierbein fields $E^\mu{}_A$:
\begin{subequations}\label{inverse}
\begin{eqnarray}
{\hat{E}}^\mu{}_A &=& \frac{1}{c} \, \tau^\mu{}_A  - \frac{1}{c^3} \, \tau{}^\mu{}_B \, \bigl( m_\nu{}^B - C_\nu{}^B \bigr) \, \tau^\nu{}_A\ + \ O(c^{-5})\,,\\[.2truecm]
\hat{E}^\mu{}_{A^\prime}{} &=& E^\mu{}_{A^\prime} - \frac{1}{c^2} \, \tau^\mu{}_A \, \bigl( m_\nu{}^A - C_\nu{}^A \bigr) \, E^\nu{}_{A^\prime}\ + \ O(c^{-4})\,.
\end{eqnarray}
\end{subequations}

Plugging \eqref{eq:Vielbeinexp} and \eqref{eq:Omegaexp} into the torsionlessness constraint \eqref{eq:torsionless}, we derive the following expansions in $c$\,:
\begin{subequations} \label{eq:limitconcon}
\begin{align}
	c \, R_{\mu\nu}{}^A (H) + \frac{1}{c} \bigl( R_{\mu\nu}{}^A (Z) - 2 \, C_{\mu\nu}{}^A \bigr) + O(c^{-3}) = 0\,, \label{eq:cR+} \\
	R_{\mu\nu}{}^{A'} \! (P) + O (c^{-2}) = 0\,,
\end{align}
\end{subequations}
where $R_{\mu\nu}{}^A (H)$\,, $R_{\mu\nu}{}^{A'}(P)$ and $R_{\mu\nu}{}^A (Z)$ are associated with the longitudinal (resp. transverse) translational generator $H_A$ (resp. $P_{A'}$) and a noncentral extension in the string Newton-Cartan algebra, respectively. They are formally defined as in the first three equations of \eqref{eq:curvaturetwoforms}, which we repeat here,
\begin{subequations}
\begin{align}
	R_{\mu\nu}{}^A (H) & = 2 \bigl( \p^{}_{[\mu} \tau^{}_{\nu]}{}^{A} + \epsilon^A{}_{B} \tau^{}_{[\mu}{}^B \Omega^{}_{\nu]} \bigr) \,, \\[4pt]
	R_{\mu\nu}{}^{A'} (P) & = 2 \bigl( \p^{}_{[\mu} E^{}_{\nu]}{}^{A'} + E^{}_{[\mu}{}^{B'} \Omega^{}_{\nu]}{}^{A'}{}_{B'} - \tau^{}_{[\mu}{}^A \Omega^{}_{\nu]A}{}^{A'} \bigr)\,, \\[4pt]
	R_{\mu\nu}{}^A (Z) & = 2 \bigl( \p^{}_{[\mu} m^{}_{\nu]}{}^A + \epsilon^A{}_B m^{}_{[\mu}{}^B \Omega^{}_{\nu]} + \tau^{}_{[\mu}{}^B n^{}_{\nu]}{\epsilon}^A{}_B + E^{}_{[\mu}{}^{A'} \Omega^{}_{\nu]}{}^A{}_{A'}  \bigr)\,. \end{align}
\end{subequations}
We have also defined
\be \label{eq:CmunuA}
	C_{\mu\nu}{}^A \equiv \p_{[\mu} C_{\nu]}{}^A + \epsilon^A{}_B \, C_{[\mu}{}^B \Omega_{\nu]}\,.
\ee
Furthermore, from the zero curvature constraint $\p_{[\mu} \hat{M}_{\nu\rho]} = 0$ in \eqref{constraint}, we obtain
\begin{align} \label{eq:XX0}
	0 & = X_{[\mu}{}^A \p^{}_\nu X_{\rho]}{}^B \epsilon_{AB} \notag \\[2pt]
		& = \tfrac{1}{2} \epsilon_{AB} \Bigl\{ c^2 \, \tau_{[\mu}{}^A R_{\nu\rho]}{}^B (H) - \ls C_{[\mu}{}^A R_{\nu\rho]}{}^B (H) + 2 \tau_{[\mu}{}^A C_{\nu\rho]}{}^B \rs \Bigr\} + O(c^{-2})\,.
\end{align}
Using \eqref{eq:cR+} to solve for $R_{\mu\nu}{}^A (H)$ in \eqref{eq:XX0} and then taking the $c \rightarrow \infty$ limit, we obtain\,\footnote{Note that the limiting procedure does not allow us to set the full curvature $R_{\mu\nu}{}^A(Z)$ equal to zero.}
\be \label{eq:etauRZ}
	\epsilon_{AB} \, \tau_{[\mu}{}^A R_{\nu\rho]}{}^B (Z) = 0\,.
\ee
In components, \eqref{eq:etauRZ} is equivalent to the equations
\be \label{eq:limitRmnZ}
	R_{A'A}{}^A (Z) = R_{A'B'}{}^A (Z) = 0\,,
\ee
Here and in the following, we use the nonrelativistic analogue of turning curved indices $\mu$ into flat indices $A$, $A^\prime$, by which one replaces a curved spacetime index $\mu$ that is contracted with the curved spacetime index of an inverse Vielbein field by the flat index of this inverse Vielbein field. For example,
\begin{equation} \label{eq:convconst}
  R_{A^\prime A}{}^B (Z) \equiv E^{\mu}{}_{A^\prime} \tau^{\nu}{}_{A} R_{\mu\nu}{}^B (Z)\,.
\end{equation}
Taking the $c \rightarrow \infty$ limit also in \eqref{eq:limitconcon}, we obtain
\be \label{eq:RHRP}
	R_{\mu\nu}{}^A (H) = R_{\mu\nu}{}^{A'} (P) = 0\,.
\ee
Note that neither of the constraints \eqref{eq:limitRmnZ} and \eqref{eq:RHRP} depends on the gauge field $n_\mu$ nor the functions $C_\mu{}^A$\,. Using these constraints, we can solve for all the components in the spin connections $\Omega_\mu$\,, $\Omega_{\mu}{}^{A'B'}$ as well as $\tilde{\Omega}_\mu{}^{AA'}$\,, with
\begin{align}\label{omegatilde}
	\tilde{\Omega}_\mu{}^{AA'} & \equiv {\Omega}_\mu{}^{AA'} - \tau_\mu{}^B W_{B}{}^{AA'}\,.
\end{align}
Here, $W_{AB}{}^{A^\prime}$ correspond to components of $\Omega_\mu{}^{A A^\prime}$ that remain unconstrained and independent\footnote{See similar discussions in \cite{Bergshoeff:2018vfn}, in which a closely related $c \rightarrow \infty$ limit is applied to the first-order Einstein-Hilbert action in four dimensions to derive an action for the extended string Newton-Cartan gravity. After taking the limit and going to the second-order formalism, it is found that $W_{AB}{}^{A'}$ remains independent while all other components of the spin connections become dependent.}. Explicitly, they correspond to the following part of $\Omega_\mu{}^{A A^\prime}$:
\begin{align}
	W_{AB}{}^{A'} & \equiv \Omega_{(AB)}{}^{A'} - \frac{1}{2} \, \eta_{AB} \, \Omega_{C}{}^{CA'}
		= {R}_{A'(AB)} (Z) - 2 \, \bigl[ m^{}_{A'(AB)} -\tfrac{1}{2} \eta_{AB} m^{}_{A'C}{}^C \bigr]  \,. \label{eq:WABA'}
\end{align}
In terms of the independent fields $\tau_\mu{}^A$\,, $E_\mu{}^{A'}$\,, $m_\mu{}^A$ and $W_{AB}{}^{A'}$\,, we find that
\begin{subequations} \label{eq:exspinconn}
\begin{align}
	\Omega_\mu & = \epsilon^{AB} \bigl( \tau^{}_{\mu AB} - \tfrac{1}{2} \tau^{}_\mu{}^C \tau^{}_{ABC} \bigr)\,, \label{eq:exspinconnOmegamu} \\[4pt]
	\Omega_\mu{}^{A'B'} & = - 2 E^{}_{\mu}{}^{[A'B']} + E^{}_{\mu}{}^{C'} E^{A'B'}{}_{C'}  + \tau^{}_\mu{}^A \, m^{A'B'}{}_A\,, \\[4pt]
	\Omega_\mu{}^{AA'}
		& = - E^{}_\mu{}^{AA'} + E^{}_{\mu B'} E^{AA'\!B'} + m_\mu{}^{A'\!A} + \tau^{}_{\mu B} m^{AA'\!B} \notag \\[2pt]
		& \hspace{1.85cm} + 2 \tau_{\mu B} \bigl[ m^{A'(AB)} - \tfrac{1}{2} \eta^{AB} m^{A'C}{}_C \bigr] + \tau_\mu{}^B W_{B}{}^{AA'}\,. \label{eq:OmegamuAA'}
\end{align}
\end{subequations}
For details on how to solve the curvature constraints to obtain the results of \eqref{eq:exspinconn}, we refer to Appendix \ref{app:gaugingext} (see also Appendix \ref{app:gauging}).

 Next, we consider the $c \rightarrow \infty$ limit of the curvature two-form in \eqref{eq:curvature}. Plugging \eqref{eq:Omegaexp} into \eqref{eq:curvature} and comparing with the expansions made in \eqref{eq:Rexp}, we see that the $c \rightarrow \infty$ limit gives rise to
\begin{subequations} \label{eq:curvaturetwoformsMJG}
\begin{align}
R_{\mu\nu} (M) & = 2 \p^{}_{[\mu} \Omega^{}_{\nu]}\,, \\[4pt]
	R_{\mu\nu}{}^{A'B'} (J) & = 2 \bigl( \p^{}_{[\mu} \Omega_{\nu]}{}^{A'B'} + \Omega_{[\mu}{}^{A'C'} \Omega_{\nu]}{}^{B'}{}_{C'} \bigr)\,, \\[4pt]
	R_{\mu\nu}{}^{AA'} (G) & = 2 \bigl( \p^{}_{[\mu} \Omega^{}_{\nu]}{}^{AA'} + \epsilon^A{}_B \Omega_{[\mu}{}^{BA'} \Omega_{\nu]} + \Omega_{[\mu}{}^{AB'} \Omega_{\nu]}{}^{A'}{}_{B'} \bigr) \,.
\end{align}
\end{subequations}
These expressions are in form the same as the ones in \eqref{eq:curvaturetwoforms}. Using \eqref{eq:exspinconn}, both $R_{\mu\nu} (M)$ and $R_{\mu\nu}{}^{A'B'} (J)$ can be written in terms of $\tau_\mu{}^A$\,, $E_\mu{}^{A'}$ and $m_\mu{}^A$\,, with no dependence on $W_{ABA'}$\,; however, $R_{\mu\nu}{}^{AA'} (G)$ contains a $W_{ABA'}$ dependence, namely,
\begin{align} \label{eq:RAA'W}
	R_{\mu\nu}{}^{AA'} (G) & = 2 \, \Bigl( \p^{}_{[\mu} \tilde{\Omega}^{}_{\nu]}{}^{AA'} + \epsilon^A{}_B \tilde{\Omega}_{[\mu}{}^{BA'} \Omega_{\nu]} + \tilde{\Omega}_{[\mu}{}^{AB'} \Omega_{\nu]}{}^{A'}{}_{B'} \Bigr) \notag \\[4pt]
		& \quad - 2 \, \eta^{}_{BC} \, \tau_{[\mu}{}^B \lr \p_{\nu]} W{}^{ACA'} - \Omega_{\nu]}{}^{A'}{}_{B'} W{}^{ACB'} + 2 \, \Omega_{\nu]} \, \epsilon_D{}^{(A} W^{C)DA'} \rr.
\end{align}

Next, taking the large $c$ expansion of the connection in \eqref{eq:connection}, we obtain
\be \label{eq:sNCGamma0}
	\hat{\Gamma}^\rho{}_{\mu\nu} =  \Gamma^\rho{}_{\mu\nu} + O(c^{-2})\,,
\ee
with $\Gamma^\rho{}_{\mu\nu}$ defined by
\begin{align} \label{eq:sNCGamma}
  \Gamma^\rho{}_{\mu\nu} \equiv \tau^\rho{}_A \bigl[ \p_{(\mu} \tau_{\nu)}{}^A - \epsilon^{A}{}_B \Omega_{(\mu} \tau_{\nu)}{}^B \bigr]
		& + E^\rho{}_{A'} \bigl[ \p_{(\mu} E_{\nu)}{}^{A'} - \Omega_{(\mu}{}^{A'B'} E_{\nu) B'} + \tilde{\Omega}_{(\mu}{}^{AA'} \tau_{\nu) A} \bigr] \notag \\[2pt]
		& + E^\rho{}_{A'} \tau_\mu{}^A \tau_\nu{}^B W_{AB}{}^{A'}\,.
\end{align}
The associated Riemann tensor can also be derived from the large $c$ limit of \eqref{eq:Riemann},
\be \label{eq:sNCRiemann0}
	\hat{R}^\rho{}_{\sigma\mu\nu} (\hat{\Gamma}) = R^\rho{}_{\sigma\mu\nu} (\Gamma) + O (c^{-2})\,,
\ee
where
\begin{align} \label{eq:sNCRiemann}
	R^\rho{}_{\sigma\mu\nu} (\Gamma) & = \p_\mu \Gamma^\rho{}_{\sigma\nu} - \p_\nu \Gamma^\rho{}_{\sigma\mu} + \Gamma^\rho{}_{\mu\lambda} \Gamma^\lambda{}_{\sigma\nu} - \Gamma^\rho{}_{\nu\lambda} \Gamma^\lambda{}_{\mu\sigma} \notag \\
		& = - \tau^\rho{}_A \tau_\sigma{}^B R_{\mu\nu}{}^{A}{}_{B} (M) + E^\rho{}_{A'} \tau_\sigma{}^A R_{\mu\nu A}{}^{A'} (G) - E^\rho{}_{A'} E_{\sigma}{}^{B'} R_{\mu\nu}{}^{A'}{}_{B'} (J)\,.
\end{align}
which contains the following dependence on $W_{ABA'}$\,:
\be \label{eq:sNCRiemannW}
	R^\rho{}_{\sigma\mu\nu} (\Gamma) \supset - 2 E^\rho{}_{A'} \tau_\sigma{}^A \tau_{[\mu}{}^B \bigl( \p_{\nu]} W_{AB}{}^{A'} - \Omega_{\nu]}{}^{A'B'} W_{ABB'} + 2 \Omega_{\nu]} \, \epsilon^C{}_{(A} W_{B)C}{}^{A'} \bigr)\,.
\ee

\subsection{Nonrelativistic Gauge Symmetries} \label{sec:gssNC}

The various string Newton-Cartan gauge fields that appeared in the limit of the relativistic Vielbein and spin connection transform under gauge symmetries. These gauge transformations can also be obtained as a limit. In order to do this, we define the following expansions of the relativistic parameters $\hat{\Lambda}_{\hat{A}\hat{B}}$ and $\hat{\eta}_\mu$:
\begin{subequations} \label{RNC4}
\begin{align}
	\hat{\Lambda}_{{A}{B}} & = \Bigl( \Lambda + \frac{1}{c^2} \, \sigma \Bigr) \epsilon_{AB}\,,
		& %
	\hat{\Lambda}_{AA^\prime} & = \frac{1}{c} \, \Lambda_{AA^\prime}\,, \\[2pt]
	\hat{\Lambda}_{A^\prime B^\prime} & = \Lambda_{A^\prime B^\prime} \,,
		& %
	 \hat{\eta}_\mu & = -\frac{1}{c}X_\mu{}^A\sigma^B\epsilon_{AB}\,.
	 \label{RNC4spec}
\end{align} 
\end{subequations}
The parameters $\Lambda$, $\Lambda_{A A^\prime}$, $\Lambda_{A^\prime B^\prime}$, $\sigma_A$ that appear in these expansions will then correspond to part of the parameters of the gauge transformations of the string Newton-Cartan fields, as we will now show. We also introduced $\sigma$ to parametrize the $O(c^{-2})$ term in the expansion of $\hat{\Lambda}_{AB}$\,, which will drop out after we take the $c \rightarrow \infty$ limit.

From \eqref{eq:Vielbeinexp}, \eqref{eq:Omegaexp} and \eqref{RNC4}, it follows that
\begin{subequations}
\begin{align}
    	\delta X_\mu{}^A + \frac{1}{c} \, \delta m_\mu{}^A
    	& = \Lambda \, \epsilon^A{}_B X_\mu{}^B \notag \\[2pt]
    	& \quad + \frac{1}{c} \bigl(
		 \Lambda \, \epsilon^A{}_B \, m_\mu{}^B + \Lambda^{AA'} E_{\mu A'} + \epsilon^A{}_B \, \tau_\mu{}^B \sigma \bigr) + O(c^{-3})\,, \label{eq:taumtrnsf} \\[.3truecm]
    \delta E_\mu{}^{A'} & =
	 - \Lambda_A{}^{A'} \tau_\mu{}^A + \Lambda^{A'}{}_{B'} E_\mu{}^{B'} + O(c^{-2}) \,, \label{eq:Emutrnsf} \\[.3truecm]
    \epsilon^{}_{AB} X_{[\mu}{}^A \delta X_{\nu]}{}^B & =
    	- \epsilon^{}_{AB} \, \tau_{[\mu}{}^A \lr \p_{\nu]} \sigma^B - \Omega_{\nu]} \epsilon^B{}_C \, \sigma^C \rr + O(c^{-2}) \,, \label{eq:tautrnsf}\\[.3truecm]
      \delta \Omega_\mu{}^{A'B'} & =  \p_\mu \Lambda^{A'B'} + 2 \Lambda^{C'[A'} \Omega_\mu{}^{B']}{}_{C'} + O(c^{-2})\,,\label{omegamu}\\[.3truecm]
      \delta \Omega_\mu + \frac{1}{c^2} \delta n_\mu & = \p_\mu \Lambda + \frac{1}{c^2} \bigl( \p_\mu \sigma + \epsilon_{AB} \Lambda^{AA'} \Omega_\mu{}^B{}_{A'} \bigr) + O(c^{-4})\,, \\[.3truecm]
  \frac{1}{c} \, \delta \Omega_\mu{}^{AA'} & = \frac{1}{c} \bigl( \p_\mu \Lambda^{AA'} + \Lambda \, \epsilon^A{}_B \Omega_\mu{}^{BA'} - \epsilon^A{}_B \Lambda^{BA'} \Omega_\mu \notag \\[2pt]
	& \qquad\qquad\qquad\!\! + \Lambda^{A'}{}_{B'} \Omega_\mu{}^{AB'} - \Lambda^{A}{}_{B'} \Omega_\mu{}^{A'B'} \bigr) + O(c^{-3})\,.
\end{align}
\end{subequations}
In the $c \rightarrow \infty$ limit, we obtain\,\footnote{In \cite{Bergshoeff:2018vfn} a closely related $c \rightarrow \infty$ limit has been applied to derive gauge transformations in extended string Newton-Cartan gravity by matching coefficients in front of $c^n$\,, $n \in \mathbb{Z}$ on both sides of the equations. It has been shown that these transformations agree with the ones derived from gauging the extended string Newton-Cartan algebra. In this paper, however, we do not assume the additional information that different orders in $c$ have to match separately. This explains why we are only able to derive a smaller collection of gauge transformations here.}
\begin{subequations} \label{eq:trnsfindep}
\begin{align}
	\delta \tau_\mu{}^A & = \Lambda \, \epsilon^A{}_B \, \tau_\mu{}^B\,,
		\qquad
	\delta E_\mu{}^{A'} = - \Lambda_A{}^{A'} \tau_\mu{}^A + \Lambda^{A'}{}_{B'} E_\mu{}^{B'}\,,\label{eq:deltatauEPoly} \\[4pt]
	\tau^\mu{}_A \delta m_\mu{}^A & = \p_A \sigma^A - \epsilon^A{}_B \sigma^B \Omega_A + \Lambda \, \epsilon^A{}_B m_A{}^B \,, \label{eq:deltam1} \\[4pt]
	E^\mu{}_{\Ap} \delta m_{\mu}{}^A & = \p_{A'} \sigma^A - \epsilon^A{}_B \sigma^B \Omega_{A'} + \Lambda \, \epsilon^A{}_B m_{A'}{}^B + \Lambda^{A}{}_{A'}\,, \label{eq:deltam2}
\end{align}
and
\begin{align}
	\delta \Omega_\mu & = \p_\mu \Lambda\,,
		\qquad%
	\delta \Omega_\mu{}^{A'B'} = \p_\mu \Lambda^{A'B'} + \Lambda^{C'A'} \Omega_\mu{}^{B'}{}_{C'} - \Lambda^{C'B'} \Omega_\mu{}^{A'}{}_{C'}\,, \\[4pt]
	\delta \Omega_\mu{}^{AA'} & = \p_\mu \Lambda^{AA'} + \Lambda \, \epsilon^A{}_B \Omega_\mu{}^{BA'} - \epsilon^A{}_B \Lambda^{BA'} \Omega_\mu + \Lambda^{A'}{}_{B'} \Omega_\mu{}^{AB'} - \Lambda^{A}{}_{B'} \Omega_\mu{}^{A'B'}\,.
\end{align}
\end{subequations}

The most general expression for $\delta m_\mu{}^A$ that obeys \eqref{eq:deltam1} and \eqref{eq:deltam2} is given by
\begin{equation} \label{eq:deltamlimit}
\delta m_\mu{}^A  = \p_\mu \sigma^A - \epsilon^A{}_B \sigma^B \Omega_\mu + \Lambda \, \epsilon^A{}_B m_\mu{}^B + \Lambda^{AA'} E_{\mu A'} - \tau_\mu{}^B \sigma^A{}_B \,.
\end{equation}
Here $\sigma^{AB}$ is a traceless two-tensor ($\sigma^A{}_A=0$); it represents an ambiguity that stems from the fact that the equations \eqref{eq:deltam1} and \eqref{eq:deltam2} do not determine all components of $\delta m_\mu{}^A$. The parameters $\Lambda$\,, $\Lambda^{A'B'}$ and $\Lambda^{AA'}$ that appear in \eqref{eq:trnsfindep} and \eqref{eq:deltamlimit} are associated with the longitudinal Lorentz transformation, transverse rotation and string-Galilean boost symmetry. They parametrize infinitesimal transformations that together with those parametrized by $\sigma^A$ and $\sigma^{AB}$ form a closed algebra. This algebra can be extended with longitudinal and transverse translations to form the string Newton-Cartan algebra, whose generators we denote as follows:
\begin{subequations} \label{eq:generatorslistbulk}
\begin{align}
    	\text{longitudinal translations} \qquad & H_A \\[2pt]
    	\text{transverse translations} \qquad & P_{A'} \\[2pt]
	\text{longitudinal Lorentz rotation} \qquad & M \\[2pt]
    	\text{string-Galilean boosts} \qquad & G_{AA'} \\[2pt]
    	\text{transverse rotations} \qquad & J_{A'B'} \\[2pt]
	\text{noncentral extensions} \qquad & Z_A \,, Z_{AB} \ \mathrm{with} \ Z^A{}_A = 0 \,.
\end{align}
\end{subequations}
The algebra that we call the ``string Bargmann algebra'' (and that was called ``string Newton-Cartan algebra'' in previous literature) corresponds to the subalgebra with $Z_{AB}$ restricted to be antisymmetric, i.e. $Z_{AB} = - Z_{BA}$\,.
The commutation relations of the string Bargmann and string Newton-Cartan algebras are given in Appendices \ref{app:gauging} and \ref{app:gaugingext}. It is interesting to note that the independent and dependent fields of string Newton-Cartan geometry and the transformation rules appearing above when taking  the limit of GR can also be obtained by applying a gauging procedure to the string Newton-Cartan algebra. This gauging is described in Appendix \ref{app:gaugingext}. We can thus conclude that the spacetime symmetry algebra that underlies the above identified  string Newton-Cartan geometry is the string Newton-Cartan algebra.

\section{Nonrelativistic String Theory in a String Newton-Cartan Background} \label{sec:appl}

After having discussed string Newton-Cartan geometry in the previous section, we will now construct the nonrelativistic string action in a string Newton-Cartan background and use this to discuss the formulation and properties of nonrelativistic string theory in curved backgrounds. We will see that the string Newton-Cartan symmetries encountered in the previous section, are symmetries of the nonrelativistic string action, that act on the background fields. We will also see that the background fields that occur in the nonrelativistic string action are not uniquely determined but, instead, form equivalence classes. We will then focus on two properties of nonrelativistic string theory. First, we discuss the quantum consistency of nonrelativistic string theory defined on a string Newton-Cartan background and present the equations of motion of the resulting string Newton-Cartan gravity. This discussion corroborates the one of \cite{Weyl} where the quantum consistency conditions are derived from a microscopic worldsheet point of view. Secondly, we discuss nonrelativistic T-duality. 

\subsection{Nonrelativistic String Theory from Relativistic String Theory} \label{sec:curvedbackground}

Nonrelativistic string theory can be treated as a subtle limit of relativistic string theory, which is essentially the same as the noncommutative open string (NCOS) limit but without D-branes \cite{Gomis:2000bd, Gomis:2005pg}. In this section, we extend the previous analysis of nonrelativistic string theory in a curved background to the most general background geometry, as the result of a singular limit of relativistic string theory.

Consider the relativistic string theory action, also known as the Polyakov action, in the presence of a metric field $\hat{G}_{\mu\nu}$\,, a Kalb-Ramond field $\hat{B}_{\mu\nu}$ and a dilaton $\hat{\Phi}$\,,
\begin{align} \label{eq:relS}
	\hat{S} & = \frac{1}{4\pi\alpha'} \int d^2 \sigma \ls \sqrt{h} \, h^{\alpha\beta} \p_\alpha x^{\mu} \p_\beta x^{\nu} \hat{G}_{\mu\nu} (x) + i \epsilon^{\alpha\beta} \p_\alpha x^\mu \p_\beta x^\nu \hat{B}_{\mu\nu} (x) \rs \notag \\
		& \quad + \frac{1}{4\pi} \int d^2 \sigma \sqrt{h} \, R^{(2)} \, \hat{\Phi} (x)\,,
\end{align}
where $\alpha'$ is the Ricci slope, $\sigma^\alpha$ are coordinates on the two-dimensional worldsheet, $h_{\alpha\beta}$ is the worldsheet metric, $R^{(2)}$ is the Ricci scalar defined with respect to $h_{\alpha\beta}$ and $x^\mu (\sigma)$, $\mu = 0, 1, \cdots, d-1$ are worldsheet scalars that parametrize the spacetime coordinates. We have performed a Wick rotation so that the worldsheet is Euclidean.
The path integral corresponding to the Polyakov action is given by
\be \label{eq:pathintegralrel}
	\mathcal{Z}_\text{rel.} = \int \mathcal{D} x^\mu \sqrt{- \hat{G}} \, \exp \bigl( - \hat{S} \, \bigr) = \int \mathcal{D} x^\mu \exp \bigl( - \hat{S}_G \, \bigr)\,,
\ee
where $\hat{G} = \det \hat{G}_{\mu\nu}$ and
\be \label{eq:SG}
	\hat{S}_G \equiv S - \frac{1}{8\pi} \int d^2 \sigma \sqrt{h} \, R^{(2)} \, \ln \sqrt{-\hat{G}}\,.
\ee
Note that we used the hatted symbols to represent spacetime ingredients in relativistic string theory. We take the following expansions with respect to large $c$\,:
\begin{align} \label{eq:GBPhizeta}
	\hat{G}_{\mu\nu} = \hat{E}_\mu{}^{\hat{A}} \hat{E}_\nu{}^{\hat{B}} \eta_{\hat{A}\hat{B}}\,,
		\qquad%
	\hat{B}_{\mu\nu} = \hat{M}_{\mu\nu} + B_{\mu\nu}\,,
		\qquad%
	\hat{\Phi} = \Phi + \ln c\,,
\end{align}
where the expansions of $\hat{E}_\mu{}^{\hat{A}}$ and the zero flux two-form $\hat{M}_{\mu\nu}$ are defined in \eqref{eq:Vielbeinexp}. It follows that
\begin{subequations} \label{eq:expansionex}
\begin{align}
	\hat{G}_{\mu\nu} & = c^2 \tau_{\mu\nu} + \bigl[ H_{\mu\nu} - \bigl( \tau_\mu{}^A C_\nu{}^B + \tau_\nu{}^A C_\mu{}^B \bigr) \, \eta_{AB} \bigr]
	+ \frac{1}{c^2} \bigl( m_\mu{}^A - C_\mu{}^A \bigr) \bigl( m_\nu{}^B - C_\nu{}^B \bigr) \eta_{AB}\,, \notag \\[2pt]
	\hat{B}_{\mu\nu} & = - c^2 \tau_\mu{}^A \tau_\nu{}^B \epsilon_{AB} + \bigl[ B_{\mu\nu} + \bigl( \tau_\mu{}^A C_\nu{}^B - \tau_\nu{}^A C_\mu{}^B \bigr) \, \epsilon_{AB} \bigr] - \frac{1}{c^2} C_\mu{}^A C_\nu{}^B \epsilon_{AB}\,, \\[6pt]
	\hat{\Phi} & = \Phi + \ln c\,,
\end{align}
\end{subequations}
where
\be
	\tau_{\mu\nu} = \tau_\mu{}^A \tau_\nu{}^B \eta_{AB}\,,
		\qquad
	H_{\mu\nu} = E_\mu{}^{A'} E_\nu{}^{B'} \delta_{A'B'} + \bigl( \tau_\mu{}^A m_\nu{}^B + \tau_\nu{}^A m_\mu{}^B \bigr) \, \eta_{AB}\,.
\ee
Note that both $\tau_{\mu\nu}$ and $H_{\mu\nu}$ are invariant under the string-Galilean boosts.

We would like to consider the $c \rightarrow \infty$ limit in \eqref{eq:relS}. Before considering this singular limit, we first introduce a useful rewriting of relativistic string theory that facilitates the limiting procedure. It is convenient to introduce the worldsheet Zweibein $e_\alpha{}^a$, $a = 1, 2$ such that
\be
	h_{\alpha\beta} = e_\alpha{}^a e_\beta{}^b \delta_{ab}\,.
\ee
We also define
\be
	\qd \equiv \frac{i}{\sqrt{h}} \, \epsilon^{\alpha\beta} \overline{e}_\alpha \nabla_\beta\,,
		\qquad%
	\overline{\qd} \equiv \frac{i}{\sqrt{h}} \, \epsilon^{\alpha\beta} e_\alpha \nabla_\beta\,,
\ee
where $\nabla_\alpha$ is the covariant derivative that is compatible with $h_{\alpha\beta}$\,. Moreover, we introduced the (Euclideanized) light-cone coordinates for the flat index $a$ on the worldsheet tangent space,
\be
	e_\alpha = e_\alpha{}^1 + i e_\alpha{}^2\,,
		\qquad
	\overline{e}_\alpha = - e_\alpha{}^1 + i e_\alpha{}^2\,.
\ee
Then, the action \eqref{eq:SG} can be rewritten as
\begin{align} \label{eq:SGG}
	\hat{S}_G & = \frac{1}{4\pi\alpha'} \int d^2 \sigma \sqrt{h} \left\{ \qd x^{\mu} \, \overline{\qd} x^{\nu} \bigl( \hat{G}_{\mu\nu} + \hat{B}_{\mu\nu} \bigr) + \alpha' R^{(2)} \, \bigl[ \hat{\Phi} - \tfrac{1}{4} \ln (-\hat{G}) \big] \right\}\,.
\end{align}
We next define
\begin{subequations}
\begin{align}
	\tau_\mu & \equiv \tau_\mu{}^0 + \tau_\mu{}^1\,,
		\qquad
	 C_\mu \equiv C_\mu{}^0 + C_\mu{}^1\,, \\
	\overline{\tau}_\mu & \equiv \tau_\mu{}^0 - \tau_\mu{}^1\,,
		\qquad
	\overline{C}_\mu \equiv C_\mu{}^0 - C_\mu{}^1\,.
\end{align}
\end{subequations}
Using the expansions defined in \eqref{eq:expansionex}, we find that $\hat{S}_G$ can be further rewritten as
\begin{align} \label{eq:SGrewriting}
	\hat{S}_G & = \frac{1}{4\pi\alpha'} \int d^2 \sigma \sqrt{h} \, \Bigl\{ \mathcal{D} x^\mu \overline{\mathcal{D}} x^\nu \bigl[ H_{\mu\nu} (x) + B_{\mu\nu} (x) \bigr] + \alpha' R^{(2)} \bigl[ \Phi - \tfrac{1}{4} \ln (- c^{-4} \hat{G}) \bigr] \Bigr\} \notag \\
		& \quad + \frac{1}{4\pi\alpha'} \int d^2 \sigma \sqrt{h} \, \Bigl\{ \lambda \, \overline{\mathcal{D}} x^\mu \tau_\mu + \overline{\lambda} \, {\mathcal{D}} x^\mu \overline{\tau}_\mu + c^{-2} \bigl( \lambda \overline{\lambda} - \lambda \, \overline{\mathcal{D}} x^\mu C_\mu - \overline{\lambda} \, {\mathcal{D}} x^\mu \overline{C}_\mu \bigr) \Bigr\}\notag\\
		& \quad + \frac{c^{-2}}{4\pi\alpha'} \int d^2 \sigma \sqrt{h} \, \qd x^\mu \overline{\qd} x^\nu \bigl( m_\mu{}^A m_\nu{}^B - m_\mu{}^A C_\nu{}^B - C_\mu{}^A m_\nu{}^B \bigr) \, \eta_{AB}\,,
\end{align}
where we have introduced the Lagrange multiplier fields $\lambda$ and $\overline\lambda$.\,\footnote{Note that this rewriting can only be done {\sl after} making the expansions \eqref{eq:expansionex} in the action. The Lagrange multiplier terms cannot be added to the original Polyakov action of relativistic string theory.}
We emphasize that, until now, we have kept $c$ as a finite constant and \eqref{eq:SGrewriting} is a direct rewriting of \eqref{eq:SGG}. Therefore, \eqref{eq:SGrewriting} still describes relativistic string theory.

When taking the limit of the last term of the first line in the above action, we use Sylvester's determinant identity to show that
\begin{align} \label{eq:Gdetdet}
	G \equiv - \lim_{c \rightarrow \infty} c^{-4} \, \hat{G} & \equiv - \lim_{c \rightarrow \infty} c^{-4} \det^{(d)} \bigl[ c^2 \tau_\mu{}^A \tau_\nu{}^B \eta_{AB} + H_{\mu\nu} + O(c^{-2}) \bigr] \notag \\
		& = \lim_{c \rightarrow \infty} c^{-4} \det^{(d)} \bigl( H_{\mu\nu} \bigr) \det^{(2)} \bigl[ \eta^{AB} + c^2 \tau_\mu{}^A H^{\mu\nu} \tau_\nu{}^B + O(c^{-2}) \bigr] \notag \\
		& = \det^{(d)} \bigl( H_{\mu\nu} \bigr) \det^{(2)} \bigl( \tau_\rho{}^A H^{\rho\sigma} \tau_\sigma{}^B \bigr)\,,
\end{align}
where $H^{\mu\nu}$ is the inverse of $H_{\mu\nu}$\,. Note that $G$ is independent of $m_\mu{}^A$\,. In the $m_\mu{}^A \rightarrow 0$ limit, $H^{\mu\nu} \rightarrow \infty$ and $\det H_{\mu\nu} \rightarrow 0$\,. However, $G$ remains finite: this can be seen most straightforwardly in the flat limit by taking
\be \label{eq:flatlimit}
    \tau_\mu{}^A \rightarrow \delta_\mu^A\,,
        \qquad
    E_\mu{}^{A'} \rightarrow \delta_\mu^{A'}\,,
        \qquad
    m_\mu{}^A \rightarrow 0\,,
\ee
in which case we have $G = 1$\,.

Finally, using the identity \eqref{eq:Gdetdet} and taking the limit $c \rightarrow \infty$ in \eqref{eq:SGrewriting}, we find that nonrelativistic string theory in a curved background is described by the following nonrelativistic string action:
\begin{align} \label{eq:nraction}
	S_G & = \frac{1}{4\pi\alpha'} \int d^2 \sigma \sqrt{h} \, \Bigl[ \mathcal{D} x^\mu \overline{\mathcal{D}} x^\nu \bigl( H_{\mu\nu} + B_{\mu\nu} \bigr) + \lambda \, \overline{\mathcal{D}} x^\mu \tau_\mu + \overline{\lambda} \, {\mathcal{D}} x^\mu \overline{\tau}_\mu \Bigr] \notag \\
		& \quad + \frac{1}{4\pi} \int d^2 \sigma \sqrt{h} \, R^{(2)} \, \bigl( \Phi - \tfrac{1}{4} \ln G \bigr)\,.
\end{align}
This string action in background fields also arises from turning on the appropriate vertex operators in nonrelativistic string theory \cite{Weyl} in the flat spacetime \eqref{eq:flatlimit}.

\subsection{Symmetries in Nonrelativistic String Theory}

The nonrelativistic string action is invariant under the symmetries of the string Newton-Cartan algebra, implemented as the transformations \eqref{eq:deltatauEPoly} and \eqref{eq:deltamlimit} on the background fields $\tau_\mu{}^A$, $E_\mu{}^{A^\prime}$ and $m_\mu{}^A$\,,\,\footnote{It was also noticed in \cite{Harmark:2018cdl} that the gauge transformation parametrized by $\sigma^{AB}$ is preserved in nonrelativistic string theory in a curved background.  We would like to thank Troels Harmark and Lorenzo Menculini for pointing this out.} provided the following transformation rule is assigned to $\lambda$ and $\overline{\lambda}$\,:
\begin{subequations} \label{eq:deltalambda2}
\begin{align}
	\delta \lambda & =  \Lambda  \lambda +  \, \mathcal{D} x^\mu \ls (\partial_\mu - \Omega_\mu) \, \overline{\sigma} + \tau_\mu \, (\sigma_{00} + \sigma_{(01)}) \rs\,,  \\[2pt]
	\delta \overline{\lambda} & = - \Lambda \overline{\lambda} + \, \overline{\mathcal{D}} x^\mu \ls (\partial_\mu + \Omega_\mu) \, \sigma + \overline{\tau}_\mu \, (\sigma_{00} - \sigma_{(01)}) \rs\,,
\end{align}
\end{subequations}
and provided the curvature constraint $R_{\mu\nu}{}^A (H) = 0$ is imposed.

If we also allow the matter fields $B_{\mu\nu}$ and $\Phi$ to transform (other than the $U(1)$ gauge symmetry of $B_{\mu\nu}$), then there are St\"uckelberg-type symmetries given by\,\footnote{Note that {when $\overline{C} = 1 / C$\,, the St\"{u}ckelberg symmetries leave the dilaton field invariant and overlaps with the Lorentz rotation of the string Newton-Cartan algebra. 
There are also} two cases where the St\"uckelberg symmetries leave the Kalb-Ramond field invariant and overlap with one of the symmetries of the string Newton-Cartan algebra. First, for zero torsion and if $C_\mu{}^A =D_\mu\sigma^A$, the St\"uckelberg symmetries coincide, up to a gauge transformation of the $B$-field, with the $\sigma^A$ symmetry of the string Newton-Cartan algebra. Second, if $C_\mu{}^A =\sigma^A{}_B\tau_\mu{}^B$ with $\sigma^A{}_A=0$, the St\"uckelberg symmetries coincide with the $\sigma^{AB}$ symmetry of the string Newton-Cartan algebra. See also \cite{Harmark:2019upf}.}
\begin{subequations} \label{eq:CmuACtrnsf}
\begin{align}
	\lambda & \rightarrow {C}^{-1} \bigl( \lambda - \qd x^\mu \, \overline{C}_\mu \bigr)\,,
		&%
	\tau_\mu & \rightarrow C \, \tau_\mu\,,
		&%
	H_{\mu\nu} & \rightarrow H_{\mu\nu} - \bigl( C_\mu{}^A \tau_\nu{}^B + C_\nu{}^A \tau_\mu{}^B \bigr) \, \eta_{AB}\,, \\[2pt]
	\overline{\lambda} & \rightarrow {\overline{C}}^{-1} \bigl( \overline{\lambda} - \overline{\qd} x^\mu C_\mu \bigr)\,,
		&%
	\overline{\tau}_\mu & \rightarrow \overline{C} \, \overline{\tau}_\mu\,,
		&%
	B_{\mu\nu} & \rightarrow B_{\mu\nu} + \bigl( C_\mu{}^A \tau_\nu{}^B - C_\nu{}^A \tau_\mu{}^B \bigr) \, \epsilon_{AB}\,,
\end{align}
and\,\footnote{Note that the combination $\Phi - \tfrac{1}{4} \textrm{ln}\,G$ occurring in the Polyakov action \eqref{eq:nraction} is invariant under these St\"uckelberg symmetries.}
\begin{align} \label{eq:lltransPhi}
	\Phi \rightarrow \Phi + \frac{1}{2} \ln \bigl( {C} \, \overline{C} \bigr)\,.
\end{align}
\end{subequations}
Expressing $H_{\mu\nu}$ in terms of $\tau_\mu{}^A$\,, $E_\mu{}^{A'}$ and $m_\mu{}^A$\,, one finds that $m_\mu{}^A$ transforms as
\be
    m_\mu \rightarrow C^{-1} m_\mu\,,
        \qquad
    \overline{m}_\mu \rightarrow \overline{C}^{-1} \overline{m}_\mu\,,
\ee
where $m_\mu \equiv m_\mu{}^0 + m_\mu{}^1$ and $\overline{m}_\mu \equiv m_\mu{}^0 - m_\mu{}^1$\,.
Note that, as suggested by the use of the same notation, the function $C_\mu{}^A$ in \eqref{eq:CmuACtrnsf} is precisely the $C_\mu{}^A$ in \eqref{eq:expansionex}. Furthermore, the  $C \overline{C}$ in \eqref{eq:CmuACtrnsf} is taken to be $c^2$ in  \eqref{eq:expansionex}.

As we have noted above, invariance of the action \eqref{eq:nraction} requires that the curvature constraint $R_{\mu\nu}{}^{A} (H) = 0$ is imposed, which by \eqref{eq:RmunuAtauOmega} is equivalent to
\be \label{eq:dcomponents}
	\p_{[\mu} \tau_{\nu]}{}^A = \epsilon^A{}_B \Omega_{[\mu} \tau_{\nu]}{}^B\,.
\ee
This constraint is necessary for invariance under the $\sigma_A$ transformations.
While $d$ components of \eqref{eq:dcomponents} can be used to solve for $\Omega_\mu$ as in \eqref{eq:RmunuAtauOmega}, the remaining $d(d-2)$ components give rise to the geometric constraints as in \eqref{eq:geoconstraints}. These geometric constraints  can be written in a compact form as
\be \label{eq:geoconstraint}
	\epsilon_{C}{}^{(A} \tau_{[\mu}{}^{B)} \p^{\phantom{\dagger}}_{\nu} \tau_{\rho]}{}^C = 0\,.
\ee

The condition \eqref{eq:geoconstraint} imposes nontrivial constraints on $C$ and $\overline{C}$ that appear in the transformations in \eqref{eq:CmuACtrnsf}. In terms of $\tau_\mu$ and $\overline{\tau}_{\mu}$, \eqref{eq:geoconstraint} becomes
\be \label{eq:geoconstraintlightcone}
	\tau_{[\mu} \p^{}_\nu \tau_{\rho]} = \overline{\tau}_{[\mu} \p^{}_\nu \overline{\tau}_{\rho]} = \overline{\tau}_{[\mu} \p^{}_\nu \tau_{\rho]} - {\tau}_{[\mu} \p^{}_\nu \overline{\tau}_{\rho]} = 0\,.
\ee
Applying the transformation rules
\be
	\tau_\mu \rightarrow C \, \tau_\mu\,,
		\qquad
	\overline{\tau}_\mu \rightarrow \overline{C} \, \overline{\tau}_\mu\,,
\ee
and then requiring that \eqref{eq:geoconstraintlightcone} hold, we find that
\be
	\tau_{[\mu} \overline{\tau}^{}_{\nu} \p_{\rho]} \bigl( \overline{C} C \bigr) = 0\,,
\ee
which in components gives
\be
	E^\mu{}_{A'}\p_\mu (\overline{C}C)  = 0 \,.
\ee

\subsection{Spacetime Equations of Motion}

As the first application, we apply the $c \rightarrow \infty$ limit to the one-loop beta-functions in relativistic perturbative string theory. By setting the $c \rightarrow \infty$ limit of the relativistic beta-functions to zero, we expect to derive the spacetime equations of motion in nonrelativistic string theory. Recall the beta-functions in relativistic string theory \cite{Callan:1985ia},\footnote{The conventions used here follow the ones in \cite{Polchinski:1998rq}, but in a different renormalization scheme \cite{Weyl}.}
\begin{subequations} \label{eq:relbetafunctions}
\begin{align}
	{\beta}^{\hat{G}}_{\mu\nu} & = \alpha' \lr \hat{R}_{\mu\nu} + 2 \hat{\nabla}_\mu \hat{\nabla}_\nu \hat{\Phi} - \tfrac{1}{4} \hat{\mathcal{H}}_{\mu\rho\sigma} \hat{\mathcal{H}}_\nu{}^{\rho\sigma} \rr + O({\alpha'}^2)\,, \label{eq:relbetafunctionsa} \\[5pt]
	{\beta}^{\hat{B}}_{\mu\nu} & = \alpha' \lr - \tfrac{1}{2} \hat{\nabla}^\rho \hat{\mathcal{H}}_{\rho\mu\nu} + \hat{\nabla}^\rho \hat{\Phi} \, \hat{\mathcal{H}}_{\rho\mu\nu} \rr + O({\alpha'}^2)\,, \label{eq:relbetafunctionsb} \\[5pt]
	{\beta}^{\hat{F}} & = \frac{d-26}{6}
		- \alpha' \lr \hat{\nabla}_\mu \hat{\nabla}^\mu \hat{\Phi} + \tfrac{1}{4} \hat{R} - \hat{\nabla}_\mu \hat{\Phi} \, \hat{\nabla}^\mu \hat{\Phi} - \tfrac{1}{48} \hat{\mathcal{H}}_{\mu\nu\lambda} \hat{\mathcal{H}}^{\mu\nu\lambda} \rr + O({\alpha'}^2)\,,
\end{align}
\end{subequations}
where $\hat{\nabla}_\mu$ is the covariant derivative that is compatible with $\hat{G}_{\mu\nu}$\,, with the Christoffel symbol
\be \label{eq:Christoffel}
	\hat{\Gamma}^\rho{}_{\mu\nu} = \frac{1}{2} \hat{G}^{\rho\sigma} \bigl( \p_\mu \hat{G}_{\nu\sigma} + \p_\nu \hat{G}_{\mu\sigma} - \p_\sigma \hat{G}_{\mu\nu} \bigr)\,.
\ee
The (3,1) tensor $\hat{R}^\rho{}_{\sigma\mu\nu}$ is the Riemann tensor defined on the spacetime manifold $\CM$\,, $\hat{R}_{\mu\nu} \equiv \hat{R}^\rho{}_{\mu\rho\nu}$ is the associated Ricci tensor and $\hat{R} \equiv \hat{G}^{\mu\nu} \hat{R}_{\mu\nu}$ is the associated Ricci scalar. The tensor $\hat{G}^{\mu\nu}$ denotes the inverse metric. Moreover,
\begin{align}
	\hat{\mathcal{H}}_{\mu\nu\rho} \equiv \p_\mu \hat{B}_{\nu\rho} + \p_\rho \hat{B}_{\mu\nu} + \p_\nu \hat{B}_{\rho\mu}\,,
		\qquad%
	\hat{F} \equiv \hat{\Phi} - \tfrac{1}{4} \ln (-\hat{G})\,.
\end{align}
The large $c$ expansions of $\hat{\Gamma}^\rho{}_{\mu\nu}$ and $\hat{R}^\rho{}_{\sigma\mu\nu}$ were  already derived in \eqref{eq:sNCGamma0} and \eqref{eq:sNCRiemann0}, respectively, with
\begin{align}
	\hat{\Gamma}^\rho{}_{\mu\nu} = \Gamma^\rho{}_{\mu\nu} + O(c^{-2})\,,
		\qquad%
	\hat{R}^\rho{}_{\sigma\mu\nu} (\hat{\Gamma}) = {R}^\rho{}_{\sigma\mu\nu} (\Gamma) + O(c^{-2})\,,
\end{align}
where $\Gamma^\rho{}_{\mu\nu}$ and $R^\rho{}_{\sigma\mu\nu}$ define the Christoffel symbol and the Riemann curvature tensor in string Newton-Cartan geometry. The corresponding Ricci tensor is then defined by $R_{\mu\nu} = R^\rho{}_{\mu\rho\nu} (\Gamma)$\,.
We can write $\Gamma^\rho{}_{\mu\nu}$ in \eqref{eq:sNCGamma} as
\begin{align} \label{eq:ChristoffelNRS}
	\Gamma^\rho{}_{\mu\nu} & = \tfrac{1}{2} E^{\rho\sigma} \bigl( \p_\mu H_{\sigma\nu} + \p_\nu H_{\sigma\mu} - \p_\sigma H_{\mu\nu} \bigr)
		+ \tfrac{1}{2} N^{\rho\sigma} \bigl( \p_\mu \tau_{\sigma\nu} + \p_\nu \tau_{\sigma\mu} - \p_\sigma \tau_{\mu\nu} \bigr) \notag \\[2pt]
		& \quad + E^{\rho A'} \tau_\mu{}^A \tau_\nu{}^B \bigl( W_{AB \Ap} + 2 m_{A'(AB)} - \eta_{AB} m_{A'C}{}^C \bigr)\,,
\end{align}
where
\be
	E^{\rho\sigma} \equiv E^\rho{}_{A'} E^{\sigma A'}\,,
		\qquad
	N^{\rho\sigma} \equiv \tau^{\rho}{}_A \tau^{\sigma A} - 2 E^{(\rho}{}_{\Ap} \tau^{\sigma)}{}_A m_\lambda{}^A E^{\lambda\Ap}
\ee
are both invariant under the string-Galilean boost transformations.
We emphasize that there does not exist a non-degenerate spacetime metric and hence string Newton-Cartan geometry is non-Riemannian. We also have the expansion
\be
	\hat{\mathcal{H}}_{\mu\nu\rho} = \mathcal{H}_{\mu\nu\rho} + O(c^{-2})\,,
		\qquad
	\mathcal{H}_{\mu\nu\rho} = \p_\mu B_{\nu\rho} + \p_\rho B_{\mu\nu} + \p_\nu B_{\rho\mu}\,.
\ee
It is then manifest that the expressions on the right hand side of \eqref{eq:relbetafunctions} are all finite in the $c \rightarrow \infty$ limit.

Next, plugging the expansions \eqref{eq:expansionex} into the left hand side of \eqref{eq:relbetafunctions}, we obtain
\begin{subequations} \label{eq:betabetabeta}
\begin{align}
	\beta^{\hat{G}}_{\mu\nu} & = c^2 \bigl[ \, \tau_{\mu}{}^A \beta(\tau)_{\nu}{}^B + \beta(\tau)_{\mu}{}^A \tau_{\nu}{}^B \, \bigr] \eta_{AB}  \notag \\[2pt]
		& \quad + \beta^H_{\mu\nu} - \bigl[ \, \beta(\tau)_\mu{}^A C_\nu{}^B + \beta(\tau)_\nu{}^A C_\mu{}^B \, \bigr] \,\eta_{AB} + O(c^{-2})\,, \label{eq:betaG} \\[4pt]
	\beta^{\hat{B}}_{\mu\nu} & = - c^2 \bigl[ \, \tau_{\mu}{}^A \beta(\tau)_{\nu}{}^B + \beta(\tau)_{\mu}{}^A \tau_{\nu}{}^B \, \bigr] \epsilon_{AB} \notag \\[2pt]
		& \quad + \beta^B_{\mu\nu} + \bigl[ \, \beta(\tau)_\mu{}^A C_\nu{}^B - \beta(\tau)_\nu{}^A C_\mu{}^B \, \bigr] \, \epsilon_{AB} + O(c^{-2})\,, \label{eq:betaB} \\[4pt]
		\beta^{\hat{F}} & = \beta^F + O(c^{-2})\,, \qquad F \equiv \Phi - \frac{1}{4} \ln G\,, \label{eq:betaFc2}
\end{align}
\end{subequations}
with $G$ given by \eqref{eq:Gdetdet}. Note that the right hand side of the equations in \eqref{eq:relbetafunctions} are not divergent in $c$\,. Therefore, plugging \eqref{eq:betaG} and \eqref{eq:betaB} into \eqref{eq:relbetafunctionsa} and \eqref{eq:relbetafunctionsb}, in the $c \rightarrow \infty$ limit, we obtain
\begin{align} \label{eq:betatau}
	\tau^{\mu A} \beta(\tau)_\mu{}^{B} + \tau^{\mu B} \beta(\tau)_\mu{}^{A} & = E^\mu{}_{A'} \beta(\tau)_\mu{}^A = O(c^{-2})\,.
\end{align}
Due to the undetermined $O(c^{-2})$ contributions in \eqref{eq:betatau}, the $c^2$ terms in \eqref{eq:betabetabeta} will contribute at the order $O(c^{0})$\,. Using \eqref{eq:betaG} and \eqref{eq:betaB} to form combinations in which all contributions from the $c^2$ terms in \eqref{eq:betabetabeta} cancel, and then taking the $c \rightarrow \infty$ limit, we obtain,
\begin{subequations} \label{eq:betaHB}
\begin{align}
	E^\mu{}_{A'} E^\nu{}_{B'} \beta^H_{\mu\nu} & = \alpha' P_{A'B'} + O({\alpha'}^2)\,, \\[2pt]
	E^\mu{}_{A'} E^\nu{}_{B'} \beta^B_{\mu\nu} & = \alpha' Q_{A'B'} + O({\alpha'}^2) \,, \\[2pt]
	\tau^\mu{}_A \tau^\nu{}_B \bigl( \eta^{AB} \beta^H_{\mu\nu} - \epsilon^{AB} \beta^B_{\mu\nu} \bigr) & = \alpha' \bigl( \eta^{AB} P_{AB} - \epsilon^{AB} Q_{AB} \bigr) + O({\alpha'}^2)\,, \\[2pt]
	\tau^\mu{}_B E^\nu{}_{A'} \bigl( \delta_A^B \, \beta^H_{\mu\nu} + \epsilon_A{}^B \beta^B_{\mu\nu} \bigr) & = \alpha' \bigl( P_{AA'} + \epsilon_A{}^B Q_{BA'} \bigr) + O({\alpha'}^2)\,,
\end{align}
\end{subequations}
where
\begin{subequations}
\begin{align}
	P_{\mu\nu} & \equiv R_{\mu\nu} + 2 \nabla_\mu \nabla_\nu \Phi - \tfrac{1}{4} \mathcal{H}_{\mu\Ap\Bp} \mathcal{H}_{\nu}{}^{\Ap\Bp} \,, \\[3pt]
	Q_{\mu\nu} & \equiv - \tfrac{1}{2} \nabla^\Ap \mathcal{H}_{\Ap\mu\nu} + \nabla^\Ap \Phi \, \mathcal{H}_{\Ap\mu\nu}\,,
\end{align}
\end{subequations}
and $\nabla_\mu$ is defined with respect to the connection $\Gamma^\rho{}_{\mu\nu}$ in \eqref{eq:ChristoffelNRS}.  Finally, taking the $c \rightarrow \infty$ limit in \eqref{eq:betaFc2}, we obtain
\be \label{eq:tildebetaF}
	{\beta}^F = \frac{d-26}{6} - \alpha' \bigl( \nabla_\Ap \nabla^\Ap \Phi + \tfrac{1}{4} R_{A'}{}^{A'}- \nabla^\Ap \Phi \nabla_\Ap \Phi - \tfrac{1}{48} \mathcal{H}_{\Ap\Bp\Cp} \mathcal{H}^{\Ap\Bp\Cp} \bigr) + O({\alpha'}^2)\,.
\ee
Importantly, all $C_\mu{}^A$ dependences drop out after taking the $c \rightarrow \infty$ limit, and all the beta-functions are invariant under the symmetry transformations in \eqref{eq:CmuACtrnsf}. This is demonstrated explicitly in \cite{Weyl}, where the beta-functions in nonrelativistic string theory are computed by applying the worldsheet Weyl transformations to vertex operators. The worldsheet treatment in \cite{Weyl} is intrinsic to nonrelativistic string theory without referring to the parent relativistic string theory.

Setting the beta-functions in \eqref{eq:betaHB} and \eqref{eq:tildebetaF} to zero leads to the equations of motion of string Newton-Cartan gravity coupled to the $B$-field and dilaton, namely,\,\footnote{In \cite{Kluson:2019ifd}, the same limit applied to the Type IIB supergravity equations of motion has been considered, but only for fixed $C = \overline{C} = 1$ and $C_\mu{}^A = \frac{1}{2} m_\mu{}^A$. As a result, the string Newton-Cartan equations of motion considered in \cite{Kluson:2019ifd} do not respect the symmetries given in \eqref{eq:CmuACtrnsf}.}
\begin{subequations}
\begin{align}
	P_{A'B'} & = Q_{A'B'} = \eta^{AB} P_{AB} - \epsilon^{AB} Q_{AB} = P_{AA'} + \epsilon_A{}^B Q_{BA'} = 0\,, \\[2pt]	
	\frac{d-26}{6} & = \alpha' \bigl( \nabla_\Ap \nabla^\Ap \Phi + \tfrac{1}{4} R_{A'}{}^{A'}- \nabla^\Ap \Phi \nabla_\Ap \Phi - \tfrac{1}{48} \mathcal{H}_{\Ap\Bp\Cp} \mathcal{H}^{\Ap\Bp\Cp} \bigr)\,,
\end{align}
\end{subequations}
which are supplemented by the geometric constraints given in \eqref{eq:geoconstraint}. These are consistent with the results given in \cite{Weyl}.

We note one last caveat. Recall that in \S\ref{sec:limit} we found that the independent fields consist of not only $\{ \tau_\mu{}^A, E_\mu{}^{A'}, m_\mu{}^A \}$ but also one more independent gauge field $W_{ABA'}$ that is part of the spin connection $\Omega_{\mu}{}^{AA'}$\,. But $W_{ABA'}$ does not appear in the string action \eqref{eq:nraction} and hence should not show up in the beta-functions. However, as we have seen in \eqref{eq:sNCGamma} and \eqref{eq:sNCRiemannW}, ${\Gamma}^\rho{}_{\mu\nu}$ and $R_{\mu\nu}$ both contain $W_{ABA'}$ dependences,
\begin{subequations} \label{eq:GammaRW}
\begin{align}
	\Gamma^\rho{}_{\mu\nu} & \supset E^\rho{}_{A'} \tau_\mu{}^A \tau_\nu{}^B W_{AB}{}^{A'}\,, \\[2pt]
	R_{\mu\nu} (\Gamma) & \supset \tau_\mu{}^A \tau_{\nu}{}^B \bigl( \p_{A'} W_{AB}{}^{A'} - \Omega_{A'}{}^{A'B'} W_{ABB'} + 2 \Omega_{A'} \, \epsilon^C{}_{(A} W_{B)C}{}^{A'} \bigr)\,.
\end{align}
\end{subequations}
Plugging \eqref{eq:GammaRW} into \eqref{eq:betaHB}, and making use of the tracelessness condition $W_A{}^{AA'} = 0$\,, it is a straightforward calculation to show that, reassuringly, all $W_{ABA'}$ terms drop out in the expressions \eqref{eq:betaHB}.

\subsection{Nonrelativistic T-Duality from String Theory} \label{sec:TDuality}

As a second application of the limiting procedure that we developed in this paper, we consider the $c \rightarrow \infty$ limit of the Buscher rules describing  the T-duality transformations of relativistic string theory, and reproduce the longitudinal and transverse T-duality which we derived from first principles at the  level of nonrelativistic string theory in \cite{Bergshoeff:2018yvt}.

We start with the relativistic string theory action in \eqref{eq:relS} and assume that there is a spatial Killing vector $k^\mu$ in the target space. We introduce a coordinate system $x^\mu = (y, x^{i})$
adapted to $k^\mu$,  such that $k^\mu \p_\mu = \p_y$\,. Then, the associated abelian isometry is represented by a translation in the spatial direction $y$. Performing a T-duality transformation along this $y$ direction, we obtain the dual action
\begin{align}  \label{eq:dualaction}
	\tilde{S} & = \frac{1}{4\pi\alpha'} \int d^2 \sigma \ls \sqrt{h} \, h^{\alpha\beta} \p_\alpha x^{\mu} \p_\beta x^{\nu} \tilde{G}_{\mu\nu} (x) + i \epsilon^{\alpha\beta} \p_\alpha x^\mu \p_\beta x^\nu \tilde{B}_{\mu\nu} (x) \rs \notag \\
		& \quad + \frac{1}{4\pi} \int d^2 \sigma \sqrt{h} \, R^{(2)} \, \tilde{\Phi} (x)\,,
\end{align}
with the following Buscher rules \cite{Buscher:1987sk,Buscher:1987qj,RocekVerlinde}:
\begin{subequations} \label{eq:relBuscher}
\begin{align}
	\tilde{G}_{yy} & = \frac{1}{\hat{G}_{yy}}\,,
		&
	\tilde{\Phi} & = \hat{\Phi} - \frac{1}{2} \ln \hat{G}_{yy}\,, \\[2pt]
	\tilde{G}_{yi} & = \frac{\hat{B}_{yi}}{\hat{G}_{yy}}\,,
		&
	\tilde{B}_{yi} & = \frac{\hat{G}_{yi}}{\hat{G}_{yy}}\,, \\[2pt]
	\tilde{G}_{ij} & = \hat{G}_{ij} + \frac{\hat{B}_{yi} \hat{B}_{yj} - \hat{G}_{yi} \hat{G}_{yj}}{\hat{G}_{yy}}\,,
		&
	\tilde{B}_{ij} & = \hat{B}_{ij} + \frac{\hat{B}_{yi} \hat{G}_{yj} - \hat{B}_{yj} \hat{G}_{yi}}{\hat{G}_{yy}}\,,
\end{align}
\end{subequations}
where the indices $i$, $j$ indicate the directions other than the $y$ direction.
In the following, we consider the $c \rightarrow \infty$ limit of the Buscher rules \eqref{eq:relBuscher}. We will discuss all three cases studied in \cite{Bergshoeff:2018yvt}.

\begin{itemize}

\item

\textbf{Longitudinal spatial T-duality.}
One choice is to take the isometry direction $y$  to be the longitudinal spatial direction in string Newton-Cartan geometry. Then,
\begin{equation} \label{eq:longitudinalKilling}
\tlV_y{}^0 =0, \ \ \ \tlV_y{}^1 \ne 0\,,\ \ \ \ttrV_y{}^{A^\prime}= 0 \hskip .5truecm \rightarrow \hskip .5truecm \tau_y = - \bar\tau_y \ne 0\,.
\ee
Plugging the expansions in \eqref{eq:expansionex} into \eqref{eq:relBuscher} and using \eqref{eq:longitudinalKilling}, we obtain after taking the limit $c \rightarrow \infty$:
\begin{subequations} \label{eq:longtrans}
\begin{align}
	\tilde{G}_{yy} & = 0\,,
		&
	& \hspace{-6.01cm}\tilde{\Phi} = \Phi - \frac{1}{2} \ln \tl_{yy}\,, \label{eq:longtrans1}\\
	\tilde{G}_{yi} & = \frac{\tlV_i{}^A \tlV_y{}^B \epsilon_{AB}}{\tl_{yy}}\,,
		&
	& \hspace{-6.22cm}\tilde{\tB}_{yi} = \frac{\tl_{yi}}{\tl_{yy}}\,, \\[2pt]
	\tilde{G}_{ij} & = \ttr_{ij} + \frac{\lr \tB_{yi} \tlV_j{}^A + \tB_{yj} \tlV_i{}^A \rr \tlV_{y}{}^B \epsilon_{AB} + \ttr_{yy} \tl_{ij} - \ttr_{yi} \tl_{yj} - \ttr_{yj} \tl_{yi}}{\tl_{yy}}\,, \\
	\tilde{\tB}_{ij} & = \tB_{ij} + \frac{\tB_{yi} \tl_{yj} - \tB_{yj} \tl_{yi} - \lr \ttr_{yy} \tlV_i{}^A \tlV_j{}^B - \ttr_{yi} \tlV_y{}^A \tlV_{j}{}^B + \ttr_{yj} \tlV_y{}^A \tlV_{i}{}^B \rr \epsilon_{AB}}{\tl_{yy}}\,,
\end{align}
\end{subequations}
which is independent of the $C$, $\overline{C}$ and $C_\mu{}^A$ symmetry transformations \eqref{eq:CmuACtrnsf} for $C = \overline{C}$\,. Note that, in general, $C \neq \overline{C}$ \eqref{eq:CmuACtrnsf}; however, since the boost symmetry is already fixed by committing to a coordinate system adapted to the longitudinal isometry direction $y$, we can only take $C = \overline{C}$. The dual action is still \eqref{eq:dualaction}, which describes relativistic string theory with a lightlike isometry in spacetime.
The above rules precisely reproduce the Buscher rules for longitudinal T-duality transformations in nonrelativistic string theory derived in \cite{Bergshoeff:2018yvt}. 
Moreover, the T-dual of the general geometric constraints \eqref{eq:geoconstraint} implies the following additional constraints on matter coupled GR:
\begin{equation}\label{additional}
\partial_{[i}\tilde G_{j]y} =0\,,\hskip 1.5truecm \partial_{[i}\tilde B_{j]y}=0\,.
\end{equation}
We summarize the above result in the following diagram:
$$
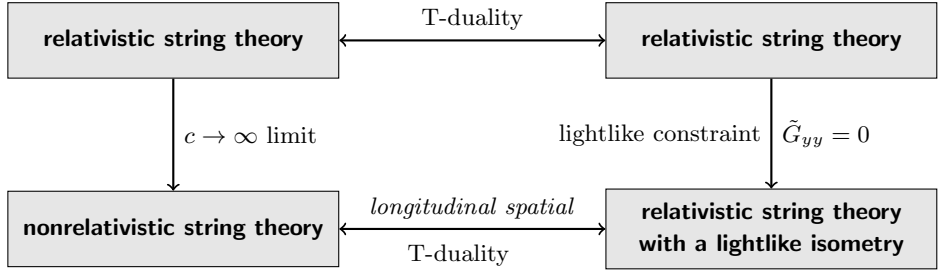
\centering
\begin{tikzpicture}
[node distance = 1cm, auto,font=\footnotesize,
every node/.style={node distance=2.5cm},
comment/.style={rectangle, inner sep= 5pt, text width=3.5cm, node distance=0.25cm, font=\scriptsize\sffamily},
force/.style={rectangle, draw, fill=black!10, inner sep=5pt, text width=4cm, text badly centered, minimum height=1cm, font=\bfseries\footnotesize\sffamily}]

\node [force] (NCstringl) {nonrelativistic string theory};
\node [force, above of=NCstringl] (Pstringl) {relativistic string theory};
\node [force, right=3.5cm of NCstringl] (NCstringr) {relativistic string theory \\ with a lightlike isometry};
\node [force, right=3.5cm of Pstringl] (Pstringr) {relativistic string theory};


\path[->,thick]
(Pstringl) edge node {$c \rightarrow \infty$ limit} (NCstringl);

\path[->,thick]
(Pstringr) edge node {$\tilde{G}_{yy} = 0$} (NCstringr);

\path[<->,thick]
(NCstringl) edge node {\emph{longitudinal spatial}} (NCstringr);

\node at (3.75,-0.35) {T-duality};

\node at (6.4,1.25) {lightlike constraint};

\path[<->,thick]
(Pstringl) edge node {T-duality} (Pstringr);

\end{tikzpicture}
$$

\qquad In \eqref{eq:longtrans}, a given Riemannian background with a lightlike isometry is mapped under T-duality to a family of distinct string Newton-Cartan backgrounds, i.e.~there exists an infinite family of solutions of the set of variables, $\{ H_{\mu\nu}, B_{\mu\nu}, \tau_\mu{}^A, \Phi \}$, that satisfies \eqref{eq:longtrans} for given $\tilde{G}_{\mu\nu}$, $\tilde{B}_{\mu\nu}$ and $\tilde{\Phi}$. This is due to the fact that the corresponding sigma model action for strings on these string Newton-Cartan backgrounds are equivalent to each other after taking into account the transformations in \eqref{eq:CmuACtrnsf}.

\qquad To prove that all solutions in \eqref{eq:longtrans} are related to each other by the transformations in \eqref{eq:CmuACtrnsf}, let us first perform the transformations \eqref{eq:CmuACtrnsf} to any given solution $\{ \tau_\mu{}^A, H_{\mu\nu}, B_{\mu\nu}, \Phi\}$ to \eqref{eq:longtrans} by fixing $\{ C = \overline{C}\,, C_y{}^1\,, C_i{}^A \}$ to be $\{ \mathscr{C}\,, \mathscr{C}_y{}^1\,, \mathscr{C}_i{}^A \}$\,, with
\begin{subequations} \label{eq:solC}
\begin{align}
    \mathscr{C} & = \frac{1}{\tau_y{}^1}\,,
        &
    \mathscr{C}_i{}^0 & \equiv \frac{1}{\tau_y{}^1} \bigl( B_{yi} - \frac{\tau_i{}^0}{2 \tau_y{}^1} H_{yy} \bigr) + \frac{\tau_i{}^1}{\tau_y{}^1} \, C_y{}^0\,, \\
    \mathscr{C}_y{}^1 & \equiv \frac{H_{yy}}{2 \tau_y{}^1}\,,
        &
    \mathscr{C}_i{}^1 & \equiv \frac{1}{\tau_y{}^1} \bigl( H_{yi} - \frac{\tau_i{}^1}{2 \tau_y{}^1} H_{yy} \bigr) + \frac{\tau_i{}^0}{\tau_y{}^1} \, C_y{}^0\,.
\end{align}
\end{subequations}
Under the above transformations, we find that $\{ \tau_\mu{}^A, H_{\mu\nu}, B_{\mu\nu}, \Phi\}$ becomes
\begin{subequations} \label{eq:primetransformation}
\begin{align}
	\tau_i^A & \rightarrow {\tau'}_i{}^A = \mathscr{C} \, \tau_i{}^A\,,
		& %
	H_{y\mu} & \rightarrow H'_{y\mu} = 0\,, \\[4pt]
	\tau_y{}^0 = 0 & \rightarrow {\tau'}_y{}^0 = 0\,,
		& %
	B_{y\mu} & \rightarrow B'_{y\mu} = 0\,, \\[4pt]
	\tau_y{}^1 & \rightarrow {\tau'}_y{}^1 = 1\,,
		&%
	H_{ij} & \rightarrow H'_{ij} = H_{ij} - (\mathscr{C}_i{}^A \tau_j{}^B + \mathscr{C}_j{}^A \tau_i{}^B) \, \eta_{AB}\,, \\[4pt]
	\Phi & \rightarrow \Phi' = \Phi + \log |\mathscr{C}|\,,
		&%
	B_{ij} & \rightarrow B'_{ij} = B_{ij} + (\mathscr{C}_i{}^A \tau_j{}^B - \mathscr{C}_j{}^A \tau_i{}^B) \, \epsilon_{AB}\,.
\end{align}
\end{subequations}

Note that $C_y{}^0$ cannot be used to fix any more degrees of freedom. Furthermore, there are in total $d^2$ degrees of freedom in $\{ \tilde{G}_{yi}\,, \tilde{G}_{ij}\,, \tilde{B}_{\mu\nu}\,, \tilde{\Phi} \}$ and $d(d+2)$ degrees of freedom in $\{ H_{\mu\nu}\,, B_{\mu\nu}\,, \tau_y{}^1\,, \tau_i{}^A\,, \Phi\}$\,. Note that $\tilde{G}_{yy} = \tau_y{}^0 = 0$\,. The latter set of variables contains $2d$ more degrees of freedom, which are fixed by the choice of $\{ C\,, C_y{}^1\,, C_i{}^A\}$ in \eqref{eq:solC}. This shows that, for any given set of $\{ H_{\mu\nu}\,, B_{\mu\nu}\,, \tau_y{}^1\,, \tau_i{}^A\,, \Phi\}$ that satisfies the Buscher rules in \eqref{eq:longtrans}, there exists a $\{ \mathscr{C}\,, \mathscr{C}_y{}^1\,, \mathscr{C}_i{}^A \}$ transformation that relates $\{ H_{\mu\nu}\,, B_{\mu\nu}\,, \tau_y{}^1\,, \tau_i{}^A\,, \Phi\}$ to $\{ H'_{\mu\nu}\,, B'_{\mu\nu}\,, \tau'_i{}^1\,, \Phi'\}$ in \eqref{eq:primetransformation}.
In terms of the background fields $\{ H'_{\mu\nu}\,, B'_{\mu\nu}\,, \tau'_i{}^1\,, \Phi'\}$, the T-duality rules \eqref{eq:longtrans} read
\begin{subequations} \label{eq:specialsol}
\begin{align}
	\tilde{G}_{yy} & = 0\,,
		&
	\tilde{G}_{yi} & = {\tau'}_i{}^0\,,
		&
	\tilde{G}_{ij} & = H'_{ij}\,, \\[2pt]
	\tilde{\tB}_{yi} & = {\tau'}_i{}^1\,,
		&
	\tilde{\tB}_{ij} & = B'_{ij}\,, 		
		&%
	\tilde{\Phi} & = \Phi'\,,
\end{align}
\end{subequations}
thus establishing a one-to-one duality map between nonrelativistic string theory with a longitudinal spatial isometry and relativistic string theory with a lightlike isometry.

\qquad Finally, to write down all solutions of $\{ H_{\mu\nu}, B_{\mu\nu}, \tau_\mu{}^A, \Phi \}$ to \eqref{eq:longtrans} in terms of the relativistic data $\tilde{G}_{\mu\nu}$\,, $\tilde{B}_{\mu\nu}$ and $\tilde{\Phi}$\,, one needs to first identify one special (but not unique) solution. Then, transforming this special solution with respect to the rules in \eqref{eq:CmuACtrnsf} with $C = \overline{C}$\,, we generate the family of all solutions to \eqref{eq:longtrans}. A simple solution to \eqref{eq:longtrans} is basically already given by \eqref{eq:specialsol}, which we write explicitly below:
\begin{subequations} \label{eq:tauprime}
\begin{align}
	{\tau'}_i{}^0 & = \tilde{G}_{yi}\,,
		&
	{\tau'}_i{}^1 & = \tilde{B}_{yi}\,,
		&
	{\tau'}_y{}^1 & = 1\,,
		&
	{\tau'}_y{}^0 & = 0\,, \\[2pt]
	H'_{ij} & = \tilde{G}_{ij}\,,
		&
	B'_{ij} & = \tilde{B}_{ij}\,.
		&
	\Phi' & = \tilde{\Phi}\,,
		&
	H'_{yi} & = H'_{yy} = B'_{yi} = 0\,.
\end{align}
\end{subequations}
Next, setting $C = \overline{C}$ in \eqref{eq:CmuACtrnsf}, we can use $\{ H'_{\mu\nu}, B'_{\mu\nu}, \tau'_\mu{}^A, \Phi' \}$ to generate any string Newton-Cartan background $\{ H_{\mu\nu}, B_{\mu\nu}, \tau_\mu{}^A, \Phi \}$ that corresponds to the same string theory action, with
\begin{subequations} \label{eq:Ctransgenerator}
\begin{align}
	\tau_\mu{}^A & = C \, {\tau'}_\mu{}^A\,,
		&%
	H_{\mu\nu} & = H'_{\mu\nu} - \bigl( C_\mu{}^A \, {\tau'}_\nu{}^B + C_\nu{}^A \, {\tau'}_\mu{}^B \bigr) \, \eta_{AB}\,, \\[2pt]
	\Phi & = \Phi' + \log | C |\,,
		&%
	B_{\mu\nu} & = B'_{\mu\nu} + \bigl( C_\mu{}^A \, {\tau'}_\nu{}^B - C_\nu{}^A \, {\tau'}_\mu{}^B \bigr) \, \epsilon_{AB}\,.
\end{align}
\end{subequations}
Plugging \eqref{eq:tauprime} into \eqref{eq:Ctransgenerator}, we arrive at the inverse Buscher rules,
\begin{subequations}
\begin{align}
    	\tau_i{}^0 & = C \, \tilde{G}_{yi}\,,
        		& %
    	\Phi & = \tilde{\Phi} + \log |C|\,,
        		\hspace{7mm}%
    	H_{yi} = C_y{}^0 \tilde{G}_{yi} - C_y{}^1 \tilde{B}_{yi} - C_i{}^1\,, \\[6pt]
    	\tau_i{}^1 & = C \, \tilde{B}_{yi}\,,
        		& %
    	H_{yy} & = - 2 \, C_y{}^1\,,
        		\hspace{14mm}%
    	B_{yi} = C_y{}^0 \tilde{B}_{yi} - C_y{}^1 \tilde{G}_{yi} - C_i{}^0\,, \\[6pt]
    	\tau_y{}^0 & = 0\,,
        		& %
	H_{ij} & = \tilde{G}_{ij} + C_i{}^0 \tilde{G}_{yj} + C_j{}^0 \tilde{G}_{yi} - C_i{}^1 \tilde{B}_{yj} - C_j{}^1 \tilde{B}_{yi} \,, \\[6pt]
    	\tau_y{}^1 & = C\,,
    		&%
    	B_{ij} & = \tilde{B}_{ij} + C_i{}^0 \tilde{B}_{yj} - C_j{}^0 \tilde{B}_{yi} - C_i{}^1 \tilde{G}_{yj} + C_j{}^1 \tilde{G}_{yi} \,.
\end{align}
\end{subequations}
Each choice of $\{ C\,, C_\mu{}^A \}$ generates an element of the equivalence class $\{ H_{\mu\nu}, B_{\mu\nu}, \tau_\mu{}^A, \Phi \}$ representing a solution to \eqref{eq:longtrans}.

\qquad It is instructive to consider the simple case that one truncates the nonrelativistic string background fields  as follows:
\be \label{eq:foliationreduction}
    \tau_i{}^1 = m_\mu{}^1 =  m_y{}^0 = B_{\mu\nu} = \Phi = 0\,,\  \ \textrm{together with}\ \
     C_\mu{}^A = 0\,,   C = 1\,,
\ee
which implies, together with the use of adapted coordinates,  that
\begin{equation}\tau_y{}^0 = E_y{}^{A'}=0\ \ \textrm{and} \ \  \tau_y{}^1=1\,.
\end{equation}
Via T-duality this leads to the  following dual relativistic string background fields:
\be
    \tilde{G}_{yy} = 0\,,
        \qquad
    \tilde{G}_{yi} = \tau_i{}^0\,,
        \qquad
    \tilde{G}_{ij} = H_{ij} = E_i{}^{A'} E_j{}^{B'} \delta_{A'B'} - \tau_i{}^0 m_j{}^0  -\tau_j{}^0 m_i{}^0 \,.
\ee
This is precisely the ``null reduction" of GR considered in \cite{Julia:1994bs, Bergshoeff:2017dqq}, where it was shown that the Ricci flat condition $\tilde{R}_{y\mu} = 0$ implies the torsionlessness condition
\be
    \p_{[i} \tau_{j]}{}^0 = 0\,.
\ee
Here, $\tilde{R}_{\mu\nu}$ denotes the Ricci tensor with respective to the Lorentzian metric $\tilde{G}_{\mu\nu}$. Given the above truncations, the same geometric constraints can also be derived by performing a Kaluza-Klein reduction over the spatial direction $y$ of the string Newton-Cartan geometric constraints \eqref{eq:geoconstraint}. This is related to the fact that, given the above truncations, a Kaluza-Klein reduction over the spatial isometry direction $y$ of string Newton-Cartan gravity leads to the same Newton-Cartan gravity theory that results from a null-reduction of general relativity.
We have not checked all the details but we expect that the same remains true for the full theory in the absence of the above truncations. The only difference is that  in that case one ends up with a matter-coupled Newton-Cartan gravity theory with non-zero torsion. From the GR point of view, the non-zero torsion  is due to the fact that one is now dealing with matter coupled GR supplemented with the constraints in \eqref{additional}. From the string Newton-Cartan point of view, this is related to the fact that the general Kaluza-Klein reduction of the geometric constraints  \eqref{eq:geoconstraint} is different in the absence of truncations.

\item \textbf{Longitudinal lightlike T-duality.}
Next, if we assume that the isometry direction $y$  is a lightlike direction in string Newton-Cartan geometry, then,
\begin{equation}
\tlV_y \ne 0\,,\hskip 1truecm \bar{\tlV}_y = 0\,,\hskip 1truecm \ttrV_y{}^{A^\prime}= 0\,,
\end{equation}
which in turn gives $\hat{G}_{\mu\nu} = 0$\,. In this case, in the $c \rightarrow \infty$ limit, the relativistic Buscher rules in \eqref{eq:relBuscher} become singular. However, working at the level of nonrelativistic string theory, one can still derive the appropriate nonrelativistic Buscher rules as in \cite{Bergshoeff:2018yvt}. In this case, nonrelativistic string theory on a string Newton-Cartan background is mapped to nonrelativistic string theory on a T-dual string Newton-Cartan background with a longitudinal lightlike isometry, which is summarized in the following diagram:
$$
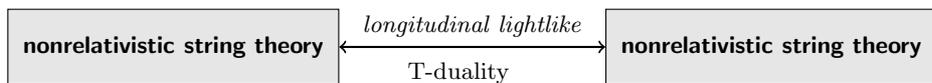
\centering
\begin{tikzpicture}
[node distance = 1cm, auto,font=\footnotesize,
every node/.style={node distance=2.5cm},
comment/.style={rectangle, inner sep= 5pt, text width=3.5cm, node distance=0.25cm, font=\scriptsize\sffamily},
force/.style={rectangle, draw, fill=black!10, inner sep=5pt, text width=4cm, text badly centered, minimum height=1cm, font=\bfseries\footnotesize\sffamily}]

\node [force] (NCstringl) {nonrelativistic string theory};
\node [force, right=3.5cm of NCstringl] (NCstringr) {nonrelativistic string theory};


\path[<->,thick]
(NCstringl) edge node {\emph{longitudinal lightlike}} (NCstringr);

\node at (3.75,-0.35) {T-duality};

\end{tikzpicture}
$$

\item \textbf{Transverse spatial T-duality.}
Finally, we assume that the isometry direction $y$  is a transverse spatial direction in string Newton-Cartan geometry. Then,
\begin{equation} \label{eq:transKilling}
	\tlV_y{}^A=0\,,\ \ \ \ttrV_y{}^{A^\prime}\ne 0\hskip .5truecm \rightarrow \hskip .5truecm \tlV_y = \bar\tlV_y = 0,\ \ \  \ttr_{yy}\ne 0\,.
 \end{equation}
Plugging the expansions  \eqref{eq:expansionex} into \eqref{eq:relBuscher} and using \eqref{eq:transKilling}, we obtain in the large $c$ expansion
\begin{subequations} \label{eq:transb}
\begin{align}
	\tilde{G}_{yy} & = \frac{1}{H_{yy}}\,,
		\qquad\quad\hspace{4cm}
	\tilde{\Phi} = \Phi + \ln c - \frac{1}{2} \ln H_{yy}\,, \\[4pt]
	\tilde{G}_{yi} & = \frac{B_{yi}}{H_{yy}}\,,
		\qquad\quad\hspace{3.75cm}
	\tilde{B}_{yi} = \frac{H_{yi}}{H_{yy}}\,, \\[4pt]
	\tilde{G}_{ij} & = c^2 \tau_{ij} + {H}_{ij} - \bigl( \tau_i{}^A C_j{}^B + \tau_j{}^A C_i{}^B \bigr) \eta_{AB} + \frac{B_{yi} B_{yj} - H_{yi} H_{yj}}{H_{yy}}\,, \\[4pt]
	\tilde{B}_{ij} & = - c^2 \tau_i{}^A \tau_j{}^B \epsilon_{AB} + B_{ij} + \bigl( \tau_i{}^A C_j{}^B - \tau_j{}^A C_i{}^B \bigr) \epsilon_{AB} + \frac{B_{yi} H_{yj} - B_{yj} H_{yi}}{H_{yy}}\,.
\end{align}
\end{subequations}
We omitted all terms of order $O(c^{-2})$\,.
Plugging the expressions \eqref{eq:transb} into the action  \eqref{eq:dualaction} and taking the $c \rightarrow \infty$ limit in the way described in \S\ref{sec:curvedbackground}, we obtain the dual action
\begin{align*} \label{eq:NCstringtransdual}
	\tilde{S}_\text{trans.} & = \frac{1}{4\pi\alpha'} \int d^2 \sigma \sqrt{h} \, \Bigl[ \mathcal{D} x^\mu \overline{\mathcal{D}} x^\nu \bigl( \tilde{H}_{\mu\nu} + \tilde{B}'_{\mu\nu} \bigr) + \lambda \, \overline{\mathcal{D}} x^\mu \tau_\mu + \overline{\lambda} \, {\mathcal{D}} x^\mu \overline{\tau}_\mu + \alpha' R^{(2)} \, \tilde{\Phi} \Bigr]\,,
\end{align*}
where
\begin{subequations} \label{eq:transBuscher}
\begin{align}
	\tilde{\ttr}_{yy} & = \frac{1}{\ttr_{yy}}\,,
		&
	\tilde{\Phi} & = \Phi - \frac{1}{2} \log \ttr_{yy}\,, \\[2pt]
	\tilde{\ttr}_{yi} & = \frac{\tB_{yi}}{\ttr_{yy}}\,,
		&
	\tilde{\tB}'_{yi} & = \frac{\ttr_{yi}}{\ttr_{yy}}\,, \\[2pt]
	\tilde{\ttr}_{ij} & = \ttr_{ij} + \frac{\tB_{yi} \tB_{yj} - \ttr_{yi} \ttr_{yj}}{\ttr_{yy}}\,,
		&
	\tilde{\tB}'_{ij} & = \tB_{ij} + \frac{\tB_{yi} \ttr_{yj} - \tB_{yj} \ttr_{yi}}{\ttr_{yy}}\,.
\end{align}
\end{subequations}
Now, the dual action describes nonrelativistic string theory with a transverse isometry in spacetime. These precisely reproduce the Buscher rules for transverse T-duality transformations in nonrelativistic string theory as derived in \cite{Bergshoeff:2018yvt}. We summarize the above result in the following diagram:
$$
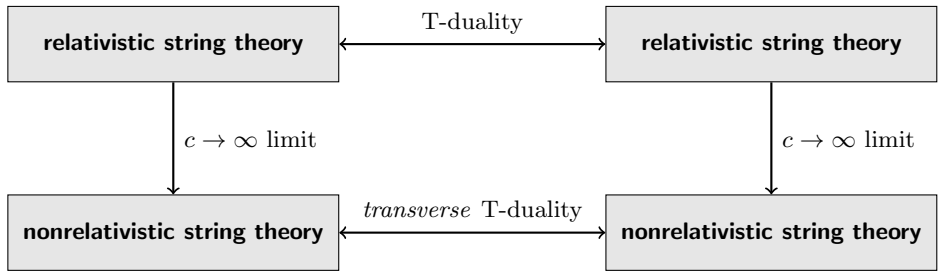
\begin{tikzpicture}
[node distance = 1cm, auto,font=\footnotesize,
every node/.style={node distance=2.5cm},
comment/.style={rectangle, inner sep= 5pt, text width=3.5cm, node distance=0.25cm, font=\scriptsize\sffamily},
force/.style={rectangle, draw, fill=black!10, inner sep=5pt, text width=4cm, text badly centered, minimum height=1cm, font=\bfseries\footnotesize\sffamily}]

\node [force] (NCstringl) {nonrelativistic string theory};
\node [force, above of=NCstringl] (Pstringl) {relativistic string theory};
\node [force, right=3.5cm of NCstringl] (NCstringr) {nonrelativistic string theory};
\node [force, right=3.5cm of Pstringl] (Pstringr) {relativistic string theory};


\path[->,thick]
(Pstringl) edge node {$c \rightarrow \infty$ limit} (NCstringl);

\path[->,thick]
(Pstringr) edge node {$c \rightarrow \infty$ limit} (NCstringr);

\path[<->,thick]
(NCstringl) edge node {\emph{transverse} T-duality} (NCstringr);

\path[<->,thick]
(Pstringl) edge node {T-duality} (Pstringr);

\end{tikzpicture}
$$

\end{itemize}

\section{Conclusions} \label{sec:conclusions}

The purpose of this work was to present an in-depth study of nonrelativistic string theory in a curved background from a spacetime point of view. The relevant geometry, replacing the Riemannian geometry underlying the relativistic string, is string Newton-Cartan geometry. One salient feature of string Newton-Cartan geometry is that it has a {\it two-dimensional} foliation dividing spacetime into two directions longitudinal to the string and the remaining spatial directions transverse to the string. The Polyakov-type action of the nonrelativistic string in such a background was obtained by taking a special nonrelativistic limit of the  Polyakov action of relativistic string theory. A noteworthy feature of this limit was that, after making a large $c$ expansion but before taking the limit $c\rightarrow\infty$\,, a rewriting of the action was required thereby introducing two Lagrange multiplier fields $\lambda$ and $\overline\lambda$. Here, we found that the nonrelativistic string action is invariant under the string Newton-Cartan algebra. This is a subalgebra of an infinite-dimensional algebra, that appears as a symmetry algebra of the nonrelativistic string action in flat spacetime. The fact that the flat space action has, unlike the relativistic case, an infinite set of symmetries is related to the fact that there is a two-dimensional foliation structure attributed to the target space. It would be interesting to further investigate the consequences of this infinite set of symmetries.

The Polyakov-type action of the nonrelativistic string can be taken as the action defining nonrelativistic string theory without the need to remember its relativistic origin.  It is the starting point of any investigation in nonrelativistic string theory. One may try to apply many of the techniques that have been applied before in a relativistic context to this nonrelativistic string theory as well. In this paper, we discussed two of them. We investigated the inverse of the nonrelativistic T-duality rules derived in \cite{Bergshoeff:2018yvt}. The definition of these inverse T-duality rules is non-trivial due to the fact that there is an equivalence relation between the string Newton-Cartan backgrounds plus the $B$-field and dilaton that couple to the nonrelativistic string. This is of special relevance when considering the longitudinal spatial T-duality that maps a string Newton-Cartan background to a general relativity background with a lightlike Killing direction. This relates nonrelativistic string theory to a wide range of topics such as the Discrete Lightcone Quantization (DLCQ) of relativistic string theory \cite{Seiberg:1997ad,Sen:1997we,Hellerman:1997yu}, nonrelativistic branes \cite{Lambert:2019} and nonrelativistic supersymmetry \cite{Silvia:2019}. This result will be important when using T-duality as a solution generating technique. In particular, it would be interesting to see what precisely happens with the T-duality between a string and wave solution in view of the fact that there are no waves in the nonrelativistic case. Last but not least, we plan to investigate what happens if one considers the longitudinal spatial T-duality in the presence of supersymmetry. Since the lightlike Killing direction is a constraint of the geometry  one should  consider the supersymmetry variations of this constraint and verify whether it leads to a finite number of further constraints.

The other application we considered was the quantum consistency of the string Newton-Cartan background. A detailed microscopic derivation from a worldsheet point of view of the vanishing beta-functions of the background fields that determine the quantum consistency, can be found in the companion paper \cite{Weyl}. In this paper, we considered the nonrelativistic limit of the one-loop beta-functions in relativistic string theory, by setting the resulting nonrelativistic beta-functions to zero we derive the same set of equations of motion of string Newton-Cartan gravity coupled to the $B$-field and dilaton as in \cite{Weyl}. It would be extremely interesting to find different solutions to these non-linear equations of motion and investigate their behaviour under nonrelativistic T-duality. Finding a solution with a non-trivial horizon could be a first step towards a nonrelativistic holography defined by the nonrelativistic string theory considered in this work.

\acknowledgments

We would like to thank Zachary Fisher, Joaquim Gomis, Kevin T. Grosvenor, Troels Harmark, Jelle Hartong, Charles Melby-Thompson, Lorenzo Menculini, Niels Obers and Jihwan Oh for useful discussions.
EB would like to thanks Tomas Ort\'\i n and Luca Romano for an early investigation of the  inverse nonrelativistic T-duality rules. EB and C\c{S} thank the ESI - University of Vienna - for its hospitality. Part of this work was initiated when EB was visiting the ESI - University of Vienna -  as a Senior Research Fellow.
ZY would like to thank the Niels Bohr Institute, the University of Groningen and Julius-Maximilians-Universit\"{a}t W\"{u}rzburg for hospitality.
The work of C\c{S} is part of the research programme of the Foundation for Fundamental Research on Matter (FOM), which is financially supported by the Netherlands Organisation for Science Research (NWO). This research was supported in part by the Perimeter Institute for Theoretical Physics. Research at Perimeter Institute is supported by the Government of Canada through the Department of Innovation, Science and Economic Development and by the Province of Ontario through the Ministry of Research, Innovation and Science.

\newpage

\appendices

In these appendices, we will discuss the different symmetry algebras that appear in nonrelativistic string theory, as well as aspects of their gauging. In Appendix \ref{app:globalsymm}, we will derive the commutation relations of the infinite-dimensional algebra discussed in \S\ref{sec:NRstring}, which is the symmetry algebra of the nonrelativistic string sigma model action in flat space. We will furthermore identify the string Bargmann algebra and the string Newton-Cartan algebra as finite-dimensional subalgebras of the infinite-dimensional algebra. Historically, the string Bargmann algebra was first considered to be the symmetry algebra of nonrelativistic string theory. Instead, as we saw in this paper, the relevant symmetry algebra is the slightly larger string Newton-Cartan algebra \cite{Harmark:2018cdl}.
In Appendix \ref{app:gauging}, we will apply a formal gauging procedure to the string Bargmann algebra. This leads to a version of string Newton-Cartan geometry that is different from the geometry that arises from the nonrelativistic limit of GR discussed in this paper. The kinematical structure of the latter geometry can be recovered by gauging the string Newton-Cartan algebra. This will be done in Appendix \ref{app:gaugingext}.

\section{Symmetry Algebras in Nonrelativistic String Theory} \label{app:globalsymm}

\subsection{Global Symmetries in Flat Spacetime}

The nonrelativistic string sigma model action in flat space has symmetries \eqref{eq:globaltrnsf}, that correspond to transverse rotations with parameter $\Lambda^{A^\prime B^\prime}$ as well as transformations that are parametrized by functions $f(X)$, $\overline{f}(\overline{X})$, $g^{A'} (X)$ and $\overline{g}^{A'} (\overline{X})$. It is useful to define the Taylor expansions of these functions as follows:
\begin{align}
	f (X) & = \sum_{n=-\infty}^{\infty} f_n X^n\,,
		&
	g^{A'} (X) = \sum_{n=-\infty}^{\infty} g^{A'}_n X^n\,, \\
	\overline{f} (\overline{X}) & = \sum_{n=-\infty}^{\infty} \overline{f}_n \overline{X}^n\,,
		&
	\overline{g}^{A'} (\overline{X}) = \sum_{n=-\infty}^{\infty} \overline{g}^{A'}_n \overline{X}^n\,.
\end{align}
The $f_n$\,, $\overline{f}_n$\,, $g^{A'}_n$\,, $\overline{g}^{A'}_n$ and $\Lambda^{A'B'}$ ($n \in \mathbb{Z}$)
then parametrize infinitesimal symmetry transformations that span an infinite-dimensional Lie algebra.

In order to write down the commutation relations of this Lie algebra, we compute the Noether charges corresponding to the transformations with parameters $f_n$\,, $\overline{f}_n$\,, $g^{A'}_n$\,, $\overline{g}^{A'}_n$ and $\Lambda^{A'B'}$. These charges are found to be:
\begin{subequations} \label{eq:infinitegenerators}
\begin{align}
	f_{n+1}: \quad M_n & = \int d\sigma^1 \, X^{n+1} \pi\,, \\
	\overline{f}_{n+1}: \quad \overline{M}_n & = \int d\sigma^1 \, \overline{X}^{n+1} \overline{\pi}\,, \\
	g^{A'}_{n+1}: \quad G_n^{A'} & = \int d\sigma^1 \Bigl( X^{n+1} \pi^{A'} - T x^{A'} \frac{\p}{\p {\sigma^1}} X^{n+1} \Bigr)\,, \\
	\overline{g}_{n+1}^{A'}: \quad \overline{G}_n^{A'} & = \int d\sigma^1 \Bigl( \overline{X}^{n+1} \pi^{A'} + T x^{A'} \frac{\p}{\p \sigma^1} \overline{X}^{n+1} \Bigr)\,, \\
	\Lambda^{A'B'}\!:\,\, J^{A'B'} & = \int d\sigma^1 \bigl( x^{A'} \pi^{B'} - x^{B'} \pi^{A'} \bigr)\,,
\end{align}
\end{subequations}
where we defined the conjugate momenta of $X$\,, $\overline{X}$ and $x^{A'}$ as follows:
\begin{align}
	\pi \equiv \tfrac{1}{2} \, T \, \lambda\,,
		\qquad%
	\overline{\pi} \equiv - \tfrac{1}{2} \, T \, \overline{\lambda}\,,
		\qquad%
	\pi^{A'} \equiv T \, \frac{\p x^{A'}}{\p \sigma^0}\,.
\end{align}
Note that
\be
	G_{-1}^{A'} = \overline{G}_{-1}^{A'} = \int d\sigma^1 \pi^{A'}\,.
\ee
Denoting equal-$\sigma^0$ Poisson brackets by $[\,.\,,.\,]$ so that
\begin{align}
	[ X (\sigma^0\,, \sigma^1), \pi (\sigma^0\,, \tilde{\sigma}^1) ] & = \delta (\sigma^1 - \tilde{\sigma}^1)\,, \\[2pt]
	[ \overline{X} (\sigma^0\,, \sigma^1) \,, \overline{\pi} (\sigma^0\,, \tilde{\sigma}^1) ] & = \delta (\sigma^1 - \tilde{\sigma}^1)\,, \\[2pt]
	[ x^{A'} (\sigma^0\,, \sigma^1), \pi^{B'} (\sigma^0\,, \tilde{\sigma}^1) ] & = \delta (\sigma^1 - \tilde{\sigma}^1) \, \delta^{A'B'}\,,
\end{align}
these Noether charges are then found to have the following non-vanishing Poisson bracket algebra ($m, n \in \mathbb{Z}$ ):
\begin{subequations} \label{eq:fullalgebra}
\begin{align}
	[M_m\,, M_n] & = (m-n) \, M_{m+n}\,,
		&
	[\overline{M}_m\,, \overline{M}_n] & = (m-n) \, \overline{M}_{m+n}\,, \\
	[G_n^{A'}\,, M_m] & = (n+1) \, G_{m+n}^{A'}\,,
		&
	[\overline{G}_n^{A'}\,, \overline{M}_m] & = (n+1) \, \overline{G}_{m+n}^{A'}\,, \\[2pt]
	[G_n^{A'}\,, J^{B'C'}] & = \delta^{A'B'} G_n^{C'} - \delta^{A'C'} G_n^{B'}\,,
		&
	[\overline{G}_n^{A'}\,, J^{B'C'}] & = \delta^{A'B'} \overline{G}_n^{C'} - \delta^{A'C'} \overline{G}_n^{B'}\,,
\intertext{and}
	[G_m^{A'}\,, G_n^{B'}] & = \delta^{A'B'} \frac{m-n}{m+n+2} \, Y_{m+n+2, 0}\,,
		&
	[Y_{m,n}\,, M_\ell] & = m Y_{m+\ell, n}\,, \\
	[\overline{G}_m^{A'}\,, \overline{G}_n^{B'}] & = \delta^{A'B'} \frac{n-m}{m+n+2} \, Y_{0, m+n+2}\,,
		& %
	[Y_{m,n}\,, \overline{M}_\ell] & = n Y_{m,n+\ell}\,, \\[2pt]
	[G_m^{A'}\,, \overline{G}_n^{B'}] & = \delta^{A'B'} Y_{m+1,n+1}\,,
\end{align}
In addition, the Poisson bracket between the $J_{A'B'}$ is
\begin{align}
	[J^{A'B'}\,, J^{C'D'}] & = \delta^{B'C'} J^{A'D'} - \delta^{A'C'} J^{B'D'} + \delta^{A'D'} J^{B'C'} - \delta^{B'D'} J^{A'C'}\,.
\end{align}
\end{subequations}
The noncentral extensions that appear in \eqref{eq:fullalgebra} are explicitly given by:
\be
	Y_{m,n} = - T \int d\sigma^1 \, \p_{\sigma^1} \bigl( X^m \overline{X}^n \bigr)\,.
\ee
For the $Y_{m,n}$, the indices $m$, $n$ are understood as  integers that are not both zero. In particular, $Y_{0,0}$ is understood to be zero, so that the above commutation relations are consistent with the transverse translations $P^{A^\prime} = G^{A^\prime}_{-1} = \overline{G}_{-1}^{A^\prime}$ commuting among themselves. These noncentral extensions $Y_{m,n}$ exist due to the fact that the Lagrangian associated with the action $S_\text{flat}$ in \eqref{eq:actionflat} is invariant with respect to the symmetries generated by $G_n^{A'}$ and $\overline{G}_n^{A'}$ up to nontrivial total derivative terms only. The above algebra contains two copies of the Witt algebra and was first introduced as  extended Galilean symmetries in \cite{Batlle:2016iel}\,\footnote{Also see the recent work \cite{FreeStringy}.}.

\subsection{The String Bargmann and Newton-Cartan Algebras as Subalgebras}

One can identify a subalgebra of the infinite-dimensional algebra \eqref{eq:fullalgebra}, associated with symmetry transformations that act on the fields of the action \eqref{eq:actionflat} as follows:
\begin{subequations}
\begin{align}
    \delta x^A & = \Xi^A - \Lambda \, \epsilon^A{}_B \, x^B\,, \\[2pt]
    \delta x^\Ap & = \Xi^\Ap \! -  \Lambda^\Ap{}_\Bp \, x^\Bp + \Lambda_{A}{}^{\Ap} x^A\,, \\[2pt]
    \delta \lambda_A & = - \Lambda \, \epsilon_A{}^B \, \lambda_B - \Lambda_{A\Ap} \, \dot{x}^\Ap \! - \epsilon_{AB} \, \Lambda^B{}_\Ap \, \p_\sigma x^\Ap\,,
\end{align}
\end{subequations}
where
\be
	\lambda_0 = \frac{1}{2} \bigl ( \lambda - \overline{\lambda} \bigr)\,,
		\qquad
	\lambda_1 = \frac{1}{2} \bigl ( \lambda + \overline{\lambda} \bigr )\,.
\ee
Here, $\Xi^A$ and $\Xi^{A'}$ parametrize the longitudinal and transverse translations, respectively, $\Lambda^{AB} \equiv \epsilon^{AB} \Lambda$ parametrizes the longitudinal Lorentz rotation, $\Lambda^{A'B'}$ parametrizes the transverse rotations and $\Lambda^{A\Ap}$ parametrizes the string-Galilean boosts. These transformations, along with extra noncentral extensions, span a Lie algebra with the following generators
\begin{subequations} \label{eq:generators}
\begin{align}
    \text{longitudinal translations} \qquad & H_A \\[2pt]
    \text{transverse translations} \qquad & P_{A'} \\[2pt]
    \text{longitudinal Lorentz rotation} \qquad & M \\[2pt]
    \text{string-Galilean boosts} \qquad & G_{AA'} \\[2pt]
    \text{transverse rotations} \qquad & J_{A'B'} \\[2pt]
    \text{noncentral extensions} \qquad & Z_A\ \textrm{and}\ Z
\end{align}
\end{subequations}
and commutation relations
\begin{subequations} \label{eq:originalalgebra}
\begin{align}
    [H_A, M] & = \epsilon_A{}^B H_B\,, &
    [H_A, G_{BA'}]  & = \eta^{}_{AB} P_{A'}\,, \\
    [P_{A'}, J_{B'C'}] & = \delta_{A'B'} P_{C'} - \delta_{A'C'} P_{B'}\,, &
    [G_{AA'}, M] & = \epsilon_A{}^B G_{BA'}\,, \\
    [G_{AA'}, J_{B'C'}] & = \delta_{A'B'} G_{AC'} - \delta_{A'C'} G_{AB'}\,, \\[4pt]
    [G_{AA'}, P_{B'}] & = \delta_{A'\!B'} Z_A\,,
        &%
   [Z_A, M] & = \epsilon_A{}^B Z_B\,, \label{eq:ESNCaGP}\\
    [G_{AA'}, G_{BB'}] & = \delta_{A'\!B'} \epsilon_{AB} Z\,,
        &%
    [H_A, Z]  & = \epsilon_{A}{}^{B} Z_B \,, \label{eq:ESNCaGG}
\end{align}
and
\begin{align}
    [J_{A'B'}, J_{C'D'}] = \delta_{B'C'} J_{A'D'} - \delta_{A'C'} J_{B'D'} + \delta_{A'D'} J_{B'C'} - \delta_{B'D'} J_{A'C'} \,.
\end{align}
\end{subequations}
This algebra is called the string Newton-Cartan algebra and was introduced in \cite{nrGalilei, stringyNC}. Its embedding in the infinite-dimensional algebra \eqref{eq:fullalgebra} is given by the following identifications:
\begin{subequations} \label{eq:algebradict}
\begin{align}
	H_0 & = M_{-1} + \overline{M}_{-1}\,,
		&
	P_{A'} & = G^{A'}_{-1} = \overline{G}^{A'}_{-1}\,, \\[4pt]
	H_1 & = M_{-1} - \overline{M}_{-1}\,,
		&
	M & = \overline{M}_0 - M_0\,, \\
	G_{0A'} & = \tfrac{1}{2} \, \bigl( G^{A'}_0 + \overline{G}_0^{A'} \bigr)\,,
		&
	Z_0 & = - \tfrac{1}{2} \, (Y_{0,1} - Y_{1,0})\,, \\[2pt]
    	G_{1A'} & = \tfrac{1}{2} \, \bigl( \overline{G}^{A'}_0 - G^{A'}_0 \bigr)\,,
		&
	Z_1 & = - \tfrac{1}{2} \, (Y_{0,1} + Y_{1,0})\,,
		&%
	Z & = \tfrac{1}{2} \, Y_{1,1}\,.
\end{align}
\end{subequations}

The string Bargmann algebra can be extended to the string Newton-Cartan algebra by replacing the generator $Z$ by a different noncentral extension $Z_{AB}$, that is a traceless two-tensor: $Z^A{}_A = 0$. The commutation relations of this extended algebra are given by the relations \eqref{eq:originalalgebra} of the string Bargmann algebra, with the last two commutation relations in eqs. \eqref{eq:ESNCaGG} replaced by
\begin{subequations}
\begin{align}
    [G_{AA'}, G_{BB'}] & = \delta_{A'\!B'} Z_{[AB]}\,, 
        &%
    [H_A, Z_{BC}]  & = 2 \eta_{AC} Z_B - \eta_{BC} Z_A\,, \\[2pt]
    & & [Z_{AB}, M] & = \epsilon_A{}^C Z_{CB} + \epsilon_B{}^C Z_{AC}\,.
\end{align} \label{eq:commPolextra}
\end{subequations}
Note that the noncentral charge $Z$ of the string Bargmann algebra corresponds to $Z \equiv -\frac12 \epsilon^{AB} Z_{AB}$ in the string Newton-Cartan algebra. The latter algebra then contains two extra noncentral charges with respect to the string Bargmann algebra. The embedding of the string Newton-Cartan algebra in the infinite-dimensional algebra is then as in \eqref{eq:algebradict}, with the addition of the following identifications for the embedding of $Z_{AB}$:
\begin{alignat}{2}
  Z_{01} &= -\frac14 \left(Y_{2,0} + Y_{0,2} \right) + \frac12 Y_{1,1} \,, \qquad & Z_{10} &= -\frac14 \left(Y_{2,0} + Y_{0,2} \right) - \frac12 Y_{1,1} \,, \nonumber \\
  Z_{00} &= Z_{11} = \frac12 \left(Y_{2,0} - Y_{0,2} \right) \,.
\end{alignat}
The string Newton-Cartan algebra appears as the local gauge symmetry algebra of nonrelativistic string theory in arbitrary backgrounds.

Note that one can extend the string Newton-Cartan algebra by a generator $D$ that acts as a dilatation, such that the following extra commutation relations are satisfied:
\begin{align} \label{eq:Dalg}
    [D\,, H_A] = H_A\,, 
		\quad
	[G_{AA'}\,, D] = G_{AA'}\,,
		\quad
	[Z_A\,, D] = Z_A\,, 
		\quad
	[Z_{AB}\,, D] = 2 Z_{AB}\,.
\end{align}  
This extended algebra is also a subalgebra of the nonrelativstic string algebra, where $D$ can be identified with $M_0 + \overline{M}_0$ in the infinite-dimensional algebra. This dilatation appears as a St\"uckelberg symmetry when formulating nonrelativistic string theory in arbitrary backgrounds in a different set of field variables. This set of variables can be found in Appendix \ref{app:gaugingext}, which differs from the ones used in the bulk of the paper.  

\section{Gauging the String Bargmann Algebra} \label{app:gauging}

In \cite{stringyNC}, it was shown that gauging the string Bargmann algebra given in \eqref{eq:originalalgebra} leads to a string Newton-Cartan geometry. In this appendix, we review and improve the discussion of this gauging procedure of the string Bargmann algebra in $d$ dimensions. This will allow us to introduce ingredients in string Newton-Cartan geometry that are useful for the bulk of this paper. At the same time, it will pave the way for the gauging of the string Newton-Cartan algebra in Appendix \ref{app:gaugingext}, that reproduces the string Newton-Cartan geometry obtained from a limit of GR.

\subsection{Transformation Rules and Curvatures}

In the gauging procedure, one first introduces a Lie algebra valued gauge field $\Theta_\mu$ that associates to each of the generators listed in \eqref{eq:generators} a corresponding gauge field as follows:
\be \label{eq:Thetamu}
	\Theta_\mu = H_A \tau_\mu{}^A + P_{A'} E_\mu{}^{A'} + G_{AA'} \Omega_\mu{}^{AA'} + M \Omega_\mu + \frac{1}{2} J_{A'B'} \Omega_\mu{}^{A'B'} + Z_A m_\mu{}^A + Z n_\mu\,.
\ee
We will ignore both the longitudinal and transverse translations parametrized by $\Xi^A$ and $\Xi^{A'}$ and instead declare that all gauge fields transform as covariant vectors under diffeomorphisms with parameter $\xi^\mu$. Once the conventional constraints have been imposed, see below, these diffeomorphisms then become equivalent to longitudinal and transverse translations (together with other gauge transformations) with parameters $\Xi^A = \tau_\mu{}^A \xi^\mu$ and $\Xi^{A'} = E_\mu{}^{A'} \xi^\mu$.
The remaining gauge transformation parameters are collected in the Lie algebra valued parameter $\Pi$ with
\be \label{eq:Pi}
	\Pi = G_{AA'} \Lambda^{AA'} + M \Lambda + \frac{1}{2} J_{A'B'} \Lambda^{A'B'} + Z_A \sigma^A +  Z \sigma\,.
\ee
Using the Lie brackets given in \eqref{eq:originalalgebra}, the gauge transformation
\be \label{eq:gaugetransf}
	\delta \Theta_\mu = \p_\mu \Pi - [\Theta_\mu, \Pi]\,,
\ee
then leads to the following transformation rules for the component gauge fields of $\Theta_\mu$:
\begin{subequations} \label{eq:trnsfindep1}
\begin{align}
	\delta \tau_\mu{}^A & = \Lambda \, \epsilon^A{}_B \, \tau_\mu{}^B\,,
		\qquad
	\delta E_\mu{}^{A'} = - \Lambda_A{}^{A'} \tau_\mu{}^A + \Lambda^{A'}{}_{B'} E_\mu{}^{B'}\,, \label{eq:rules1} \\[4pt]
	\delta m_\mu{}^A & = \p_\mu \sigma^A - \epsilon^A{}_B \sigma^B \Omega_\mu + \Lambda \, \epsilon^A{}_B m_\mu{}^B + \Lambda^{AA'} E_{\mu A'} + \epsilon^A{}_B \tau_\mu{}^B \sigma\,,  \label{eq:deltammuA}
\end{align}
and
\begin{align}
	\delta \Omega_\mu & = \p_\mu \Lambda\,,
		\qquad%
	\delta \Omega_\mu{}^{A'B'} = \p_\mu \Lambda^{A'B'} + \Lambda^{C'A'} \Omega_\mu{}^{B'}{}_{C'} - \Lambda^{C'B'} \Omega_\mu{}^{A'}{}_{C'}\,, \label{eq:conns1} \\[4pt]
	\delta \Omega_\mu{}^{AA'} & = \p_\mu \Lambda^{AA'} + \Lambda \, \epsilon^A{}_B \, \Omega_\mu{}^{B\Ap} - \epsilon^A{}_B \Lambda^{BA'} \Omega_\mu + \Lambda^{A'}{}_{B'} \Omega_\mu{}^{AB'} - \Lambda^{A}{}_{B'} \Omega_\mu{}^{A'B'}\,, \label{eq:conns2}\\[4pt]
	\delta n_\mu & = \p_\mu \sigma + \epsilon_{AB} \, \Lambda^{A}{}_{A'} \Omega_\mu{}^{BA'}\,. \label{eq:deltanmu}
\end{align}
\end{subequations}
The Vielbeine $\tau_\mu{}^A$ and $E_{\mu}{}^\Ap$ are not invertible. One can nevertheless define projective inverses $\tau^\mu{}_A$ and $E^\mu{}_\Ap$ via the following relations:
\begin{subequations}
\begin{align}
	\tau^\mu{}_A \tau_\mu{}^B & = \delta^B_A\,,
		&
	\hspace{-2cm}\tau_\mu{}^A \tau^\nu{}_A + E_\mu{}^{\Ap} E^\nu{}_{\Ap} & = \delta_\mu^\nu\,, \\[2pt]
	E_\mu{}^{\Ap} E^\mu{}_{\Bp} & = \delta^{\Ap}_{\Bp}\,,
		&
	\hspace{-2cm}\tau^\mu{}_A E_\mu{}^{\Ap} = E^\mu{}_{\Ap} \tau_\mu{}^A & = 0\,.
\end{align}
\end{subequations}
They transform as follows:
\begin{align}
\delta \tau^\mu{}_A = \Lambda \, \epsilon_A{}^B \tau^\mu{}_B + \Lambda_A{}^{B^\prime} E^\mu{}_{B^\prime}\,,
	\qquad
\delta E^\mu{}_{A^\prime} = \Lambda_{A^\prime}{}^{B^\prime} E^\mu{}_{B^\prime}\,.
\end{align}

The curvature two-form $\mathcal{F}_{\mu\nu}$ associated with $\Theta_\mu$ is
\begin{align} \label{eq:Fmunu}
	\mathcal{F}_{\mu\nu} & = \p_\mu \Theta_\nu - \p_\nu \Theta_\mu - [\Theta_\mu, \Theta_\nu] \notag \\
		& = H_A R_{\mu\nu}{}^A (H) + P_{A'} R_{\mu\nu}{}^{A'} (P) + G_{AA'} R_{\mu\nu}{}^{AA'} (G) + M R_{\mu\nu} (M) \notag \\
		& \quad + \frac{1}{2} J_{A'B'} R_{\mu\nu}{}^{A'B'} \! (J) + Z_{A} R_{\mu\nu}{}^A(Z) +  Z R_{\mu\nu}(Z)\,,
\end{align}
with the expressions for the curvature two-forms given by
\begin{subequations} \label{eq:curvaturetwoforms}
\begin{align}
	R_{\mu\nu}{}^A (H) & = 2 \bigl( \p^{}_{[\mu} \tau^{}_{\nu]}{}^{A} + \epsilon^A{}_{B} \tau^{}_{[\mu}{}^B \Omega^{}_{\nu]} \bigr) \,, \label{eq:RmunuAtauOmega} \\[4pt]
	R_{\mu\nu}{}^{A'} (P) & = 2 \bigl( \p^{}_{[\mu} E^{}_{\nu]}{}^{A'} + E^{}_{[\mu}{}^{B'} \Omega^{}_{\nu]}{}^{A'}{}_{B'} - \tau^{}_{[\mu}{}^A \Omega^{}_{\nu]A}{}^{A'} \bigr)\,, \label{eq:RmunuAtauOmega2} \\[4pt]
	R_{\mu\nu}{}^A (Z) & = 2 \bigl( \p^{}_{[\mu} m^{}_{\nu]}{}^A + \epsilon^A{}_B m^{}_{[\mu}{}^B \Omega^{}_{\nu]} + \tau^{}_{[\mu}{}^B n^{}_{\nu]}{\epsilon}^A{}_B + E^{}_{[\mu}{}^{A'} \Omega^{}_{\nu]}{}^A{}_{A'}  \bigr)\,, \label{eq:RmunuAZ} \\[4pt]
	R_{\mu\nu} (M) & = 2 \p^{}_{[\mu} \Omega^{}_{\nu]}\,, \label{eq:RM}\\[4pt]
	R_{\mu\nu}{}^{A'B'} (J) & = 2 \bigl( \p^{}_{[\mu} \Omega_{\nu]}{}^{A'B'} + \Omega_{[\mu}{}^{A'C'} \Omega_{\nu]}{}^{B'}{}_{C'} \bigr)\,, \label{eq:RJ}\\[4pt]
	R_{\mu\nu}{}^{AA'} (G) & = 2 \bigl( \p^{}_{[\mu} \Omega^{}_{\nu]}{}^{AA'} + \epsilon^A{}_B \Omega_{[\mu}{}^{BA'} \Omega_{\nu]} + \Omega_{[\mu}{}^{AB'} \Omega_{\nu]}{}^{A'}{}_{B'} \bigr) \,, \label{eq:RG}\\[4pt]
	R_{\mu\nu} (Z) & = 2 \p^{}_{[\mu} n^{}_{\nu]} - \epsilon_{AB} \, \Omega^{}_{[\mu}{}^{AA'} \Omega^{}_{\nu]}{}^{B}{}^{}_{A'}\,. \label{eq:RmunuZ}
\end{align}
\end{subequations}

\subsection{Solving the Curvature Constraints}

We next impose the curvature constraints:
\begin{align} \label{eq:threeeqs1}
	R_{\mu\nu}{}^{A} (H)
	= R_{\mu\nu}{}^{A'} (P)
	= R_{\mu\nu}{}^{A} (Z) = 0\,.
\end{align}
Note that the first condition, $R_{\mu\nu}{}^A (H) = 0$\,, imposes an analogue of the hypersurface orthogonality condition in ordinary torsionless or twistless torsional Newton-Cartan geometry. Note the difference with the curvature constraints \eqref{eq:limitRmnZ}, \eqref{eq:RHRP} that appear in the limit procedure of \S\ref{sec:limit} where not all the components of $R_{\mu\nu}{}^A (Z)$ are set to zero, as in \eqref{eq:threeeqs1}.

The transformations of $R_{\mu\nu}{}^A (H)$\,, $R_{\mu\nu}{}^\Ap (P)$ and $R_{\mu\nu}{}^A (Z)$ can be read off from the gauge transformations of the field strength $\mathcal{F}_{\mu\nu}$ defined in \eqref{eq:Fmunu},
\be
	\delta \mathcal{F}_{\mu\nu} = - [\mathcal{F}_{\mu\nu}\,, \Pi]\,,
\ee
which gives
\begin{subequations}
\begin{align}
	\delta R_{\mu\nu}{}^A (H) & = \Lambda \, \epsilon^A{}_B \, R_{\mu\nu}{}^B (H)\,, \\[2pt]
	\delta R_{\mu\nu}{}^\Ap (P) & = - \Lambda_A{}^{\Ap} R_{\mu\nu}{}^A (H) + \Lambda^{\Ap}{}_{\Bp} R_{\mu\nu}{}^\Bp (P)\,, \\[2pt]
	\delta R_{\mu\nu}{}^A (Z) & =  \Lambda \, \epsilon^A{}_B R_{\mu\nu}{}^B (Z) + \Lambda^{A}{}_{A'} R_{\mu\nu}{}^{A'} (P) \notag \\[2pt]
		& \hspace{3cm} + \sigma \,  \epsilon^A{}_B R_{\mu\nu}{}^B (H) - \epsilon^A{}_B \sigma^B R_{\mu\nu} (M)\,.
\end{align}
\end{subequations}
In order to maintain the symmetries of the theory, one needs to make sure that the constraints \eqref{eq:threeeqs1} are preserved by all symmetries. Transforming the constraints \eqref{eq:threeeqs1} under the symmetries \eqref{eq:trnsfindep1}, one finds
\be \label{eq:deltaRRR}
	\delta R_{\mu\nu}{}^A (H) = \delta R_{\mu\nu}{}^\Ap (P) = 0\,,
		\qquad
	\delta R_{\mu\nu}{}^A (Z) = - \epsilon^A{}_B \sigma^B R_{\mu\nu} (M)\,.
\ee
One thus sees that only $R_{\mu\nu}{}^A(H) = R_{\mu\nu}{}^\Ap (P) = 0$ but not $R_{\mu\nu}{}^A (Z) = 0$ are preserved by the gauge transformation rules corresponding to the algebra \eqref{eq:originalalgebra}. This can however be remedied by modifying the gauge transformation of $n_\mu$ with extra $\sigma^A$ dependent terms as in
\begin{align} \label{eq:modifieddeltanmu}
	\delta n_\mu &= \p_\mu \sigma + \epsilon_{AB} \, \Lambda^{A}{}_{A'} \, \Omega_\mu{}^{BA'} - \tfrac{1}{2} \ls \sigma^A \tau^\nu{}_A R_{\mu \nu} (M) + \tau_\mu{}^A \sigma^B \tau^\nu{}_A \tau^\rho{}_B R_{\nu\rho} (M) \rs\,.
\end{align}
In order to show that this modified transformation indeed leads to $\delta R_{\mu\nu}{}^A(Z) = 0$\,, one needs to use that
\be
\tau_{[\mu}{}^A R_{\nu\rho]}(M) = 0 \,,
\ee
which follows from the constraint $R_{\mu\nu}{}^A(H) = 0$ and its Bianchi identity.
The reason that the transformation rule of $n_\mu$ is not anymore given by the gauge algebra, is that, due to the constraint $R_{\mu\nu}{}^A (Z) = 0$ in \eqref{eq:threeeqs1}, the gauge field $n_\mu$ has become dependent. See also the discussion below.

The curvature constraints \eqref{eq:threeeqs1} constitute both proper geometric constraints as well as conventional constraints,
i.e.~constraints that can be viewed as algebraic equations that determine some fields in terms of the others. In particular, as we will now show, most of these constraints allow one to express the field $n_\mu$ and the spin connections $\Omega_\mu$\,, $\Omega_\mu{}^{A \Ap}$ and $\Omega_\mu{}^{\Ap \Bp}$ in terms of the remaining independent fields. Plugging the expressions for the dependent spin connections and $n_\mu$ into the constraints \eqref{eq:threeeqs1}, one finds that most of these are then identically satisfied, apart from a few that represent constraints on the geometry. This can be shown by considering different projections of \eqref{eq:threeeqs1} in the longitudinal Lorentzian $A$ directions and the transverse $A'$ directions.
In order to do this, it is useful to define
\be \label{eq:taumunu}
	\tau_{\mu\nu}{}^A \equiv \p_{[\mu} \tau_{\nu]}{}^A\,,
		\qquad%
	E_{\mu\nu}{}^{A'} \equiv \p_{[\mu} E_{\nu]}{}^{A'}\,,
		\qquad%
	m_{\mu\nu}{}^A \equiv \p_{[\mu} m_{\nu]}{}^A + \epsilon^A{}_B \, m_{[\mu}{}^B \Omega_{\nu]}\,.
\ee
By taking the $(A,B)$\,, $(A,\Bp)$ and $(\Ap,\Bp)$ components of equations \eqref{eq:threeeqs1}, one obtains the following equations for the components of $\Omega_\mu$\,, $\Omega_\mu{}^{A'B'}$\,, ${\Omega}_\mu{}^{AA'}$ and ${n}_\mu$\,:
%
\definecolor{DB}{rgb}{0,0.2,0.7}
\begin{table}[t!]
\centering
\begin{tabular}{|| c || c | c | c | c || c ||}
\hhline{|t:=:t:====:t:=:t|}
Total
		&$\Omega_\mu$	&$\Omega_\mu^{\phantom{\mu}A'B'}$	&${\Omega}_\mu^{\phantom{\mu}AA'}$	&${n}_\mu$				&Constraints 						\\\hhline{|:=::====::=:|}
\textcolor{DB}{$\frac{1}{2}d(d\!-\!1) (d\!+\!2)$}
		&\textcolor{DB}{$d$} 				 &\textcolor{DB}{$\frac{1}{2}d(d-2)(d-3)$}	&\textcolor{DB}{$2d(d-2)$}				&\textcolor{DB}{$d$}					&\textcolor{DB}{$d (d-2)$}				\\\hhline{|:=::====::=:|}
\multirow{5}{*}{$R_{\mu\nu}^{\phantom{\mu\nu}A} (H)$}
		&$\Omega_A$						&\multirow{2}{*}{\textcolor{DB}{0}}		&\multirow{2}{*}{\textcolor{DB}{0}} 			&\multirow{2}{*}{\textcolor{DB}{0}}		&\multirow{2}{*}{\textcolor{DB}{0}}		\\\cline{2-2}
\multirow{6}{*}{\textcolor{DB}{$d(d-1)$}}
		&\textcolor{DB}{2} 		 			& 								& 						     	  		& 		  						& 								\\\hhline{||~||----||-||}
		&$\Omega_{A'}$					&\multirow{2}{*}{\textcolor{DB}{0}}		&\multirow{2}{*}{\textcolor{DB}{0}}			&\multirow{2}{*}{\textcolor{DB}{0}}		&$\tau_{A'(AB)} = 0$ 				\\\hhline{||~||-|~~~||-||}
		&\textcolor{DB}{$d-2$} 		  		& 								&									&						     	  	&\textcolor{DB}{$3(d-2)$}				\\\hhline{||~||----||-||}
		&\multirow{2}{*}{\textcolor{DB}{0}}		&\multirow{2}{*}{\textcolor{DB}{0}}		&\multirow{2}{*}{\textcolor{DB}{0}}			&\multirow{2}{*}{\textcolor{DB}{0}}		&$\tau_{A'B'A} = 0$ 					\\\hhline{||~||~~~~||-||}
                	&                                        			&								& 									& 						     	  	&\textcolor{DB}{$(d-2)(d-3)$}			\\\hhline{|:=::====::=:|}
\multirow{5}{*}{$R_{\mu\nu}{}^{A'} \! (P)$}
		&\multirow{2}{*}{\textcolor{DB}{0}} 		&\multirow{2}{*}{\textcolor{DB}{0}}		&$\Omega_{[AB]A'}$						&\multirow{2}{*}{\textcolor{DB}{0}}		&\multirow{2}{*}{\textcolor{DB}{0}}		\\\cline{4-4}
\multirow{6}{*}{\textcolor{DB}{$\frac{1}{2}d(d-1) (d-2)$}}
		&								& 								&\textcolor{DB}{$d-2$} 					& 						     	  	& 		   						\\\hhline{||~||----||-||}
		&\multirow{2}{*}{\textcolor{DB}{0}}		&$\Omega_{AA'B'}$					&$\Omega_{(A'B')A}$					&\multirow{2}{*}{\textcolor{DB}{0}}		&\multirow{2}{*}{\textcolor{DB}{0}}		\\\cline{3-4}
 		&								&\textcolor{DB}{$(d-2)(d-3)$}  			&\textcolor{DB}{$(d-1)(d-2)$} 				&								& 		   						\\\cline{2-6}
		&\multirow{2}{*}{\textcolor{DB}{0}}		&$\Omega_{[A'B']C'}$				&\multirow{2}{*}{\textcolor{DB}{0}}			&\multirow{2}{*}{\textcolor{DB}{0}}		&\multirow{2}{*}{\textcolor{DB}{0}}		\\\cline{3-3}
		&	          			   			&\textcolor{DB}{$\frac{1}{2}(d-2)^2 (d-3)$}	& 									&								&  								\\\hhline{|:=::====::=:|}
\multirow{5}{*}{$R_{\mu\nu}^{\phantom{\mu\nu}A} (Z)$}
		&\multirow{2}{*}{\textcolor{DB}{0}}		&\multirow{2}{*}{\textcolor{DB}{0}}		&\multirow{2}{*}{\textcolor{DB}{0}}			&$n_{A}$							&\multirow{2}{*}{\textcolor{DB}{0}}		\\\hhline{||~||~|~|~|-||~||}
\multirow{6}{*}{\textcolor{DB}{$d(d-1)$}}
		&								& 								& 									&\textcolor{DB}{2} 					& 						     	 	\\\hhline{||~||----||-||}
		&\multirow{2}{*}{\textcolor{DB}{0}}		&\multirow{2}{*}{\textcolor{DB}{0}}		&$\Omega_{(AB)A'}$					&${n}_{A'}$						&\multirow{2}{*}{\textcolor{DB}{0}}		\\\hhline{||~||~|~|--||~||}
 		&								& 								&\textcolor{DB}{$3(d-2)$} 					&\textcolor{DB}{$d-2$}				& 						  		\\\hhline{||~||----||-||}
		&\multirow{2}{*}{\textcolor{DB}{0}}		&\multirow{2}{*}{\textcolor{DB}{0}}		&$\Omega_{[A'B']A}$					&\multirow{2}{*}{\textcolor{DB}{0}}		&\multirow{2}{*}{\textcolor{DB}{0}}		\\\cline{4-4}
		&	          						& 								&\textcolor{DB}{$(d-2)(d-3)$} 				&			 					&  								\\\hhline{|b:=:b:====:b:=:b|}
\end{tabular}
\caption{This table shows how to solve  for the dependent gauge fields $\Omega_\mu$\,, $\Omega_\mu{}^{A'B'}$\,, ${\Omega}_\mu{}^{AA'}$ and ${n}_\mu$ using the conventional constraints $R_{\mu\nu}{}^A (H) = R_{\mu\nu}{}^{A'}(P) = R_{\mu\nu}{}^A (Z) = 0$\,. The numbers in the table count the components in various ingredients. The decompositions of the conventional constraint equations are listed in eq.~\eqref{eq:solutionscc1}.}
\label{tb:counting1}
\end{table}
%
\begin{subequations} \label{eq:solutionscc1}
\begin{align}
	\Omega_A & =  \epsilon^{BC} \tau^{}_{CBA}\,,
		&
	\Omega_{A'} & = \epsilon^{AB} \tau^{}_{A'AB}\,, \\
	\Omega_{[AB]A'} & = E_{BAA'}\,,
		&
	\Omega_{(AB)A'} & = -2m_{A'(AB)}\,, \label{eq:OmegaABA'}\\	
	\Omega_{(A'B')A} & = -2E_{A(A'B')}\,,
		&
	\Omega_{[A'B']A} & = m_{B'A'A}\,, \\
	\Omega_{AA'B'} & = 2E_{A[B'A']}+m_{A'B'A}\,,
		&
	\Omega_{[A'B']C'} & = E_{B'A'C'}\,, \\[2pt]
	n_A & = \epsilon^{BC} \, m^{}_{CBA}\,,
		&
	n_{A'} & = \epsilon^{AB} \bigl( m_{A'AB} - \tfrac{1}{2} E_{ABA'} \bigr)\,,
\end{align}
\end{subequations}
together with the geometric constraints
\begin{align} \label{eq:geoconstraints}
	\tau^{}_{A'(AB)} = 0,
		\qquad
	\tau^{}_{A'B'A} = 0.
\end{align}
Note that only $\Omega_{[A'B']C'}$ and not $\Omega_{(A'B')C'}$ appears in \eqref{eq:solutionscc1}. The latter is determined by $\Omega_{[A'B']C'}$ via
\be
	\Omega_{(A'B')C'} = \Omega_{[C'A']B'} + \Omega_{[C'B']A'}\,.	
\ee
The equations \eqref{eq:solutionscc1} determine all components of $\Omega_\mu$\,, $\Omega_\mu{}^{A'B'}$\,, ${\Omega}_\mu{}^{AA'}$ and ${n}_\mu$. We have collected in Table \ref{tb:counting1} all information regarding which components follow from which constraints given in eq.~\eqref{eq:threeeqs1}. Explicitly, we find from \eqref{eq:solutionscc1} that
\begin{subequations} \label{eq:spinconnectionn}
\begin{align}
	\Omega^{}_\mu & = \epsilon^{AB} \bigl( \tau^{}_{\mu AB} - \tfrac{1}{2} \tau^{}_\mu{}^C \tau^{}_{ABC} \bigr)\,, \\[4pt]
	\Omega^{}_\mu{}^{A'B'} & = - 2 E^{}_{\mu}{}^{[A'B']} + E^{}_{\mu}{}^{C'} E^{A'B'}{}_{C'}  + \tau^{}_\mu{}^A \, m^{A'B'}{}_A\,, \\[4pt]
	\Omega^{}_\mu{}^{AA'} & = - E^{}_\mu{}^{AA'} + E^{}_{\mu B'} E^{AA'\!B'} + m_\mu{}^{A'\!A} + \tau^{}_{\mu B} m^{AA'\!B}\,, \\[4pt]
	n^{}_\mu{} & = \epsilon^{AB} \bigl( m^{}_{\mu AB} - \tfrac{1}{2} \tau^{}_\mu{}^C m^{}_{ABC} - \tfrac{1}{2} E^{}_{\mu}{}^{A'} E^{}_{ABA'} \bigr)\,. \label{eq:solnnmu}
\end{align}
\end{subequations}

Using \eqref{eq:spinconnectionn} for the dependent gauge fields, together with the gauge transformations of the independent fields $\tau_\mu{}^A$\,, $E_\mu{}^{A'}$ and $m_\mu{}^A$ given in \eqref{eq:trnsfindep1}, it is straightforward to show that the transformations of the spin connections, namely, $\Omega_\mu$\,, $\Omega_\mu{}^{AA'}$ and $\Omega_\mu{}^{A'B'}$\,, match the expressions in \eqref{eq:trnsfindep1}. The gauge transformation of $n_\mu$, however, gains extra $R_{\mu\nu} (M)$-dependent terms
\be
	\delta n_\mu = \p_\mu \sigma + \epsilon_{AB} \, \Lambda^{A}{}_{A'} \, \Omega_\mu{}^{BA'} - \tfrac{1}{2} \ls \sigma^A R_{\mu A} (M) + \tau_\mu{}^A \sigma^B R_{AB} (M) \rs\,,
\ee
which is precisely the transformation rule \eqref{eq:modifieddeltanmu} constructed to ensure that the curvature constraints are invariant under the gauge transformations with parameters $\sigma^A$.

%
%
\begin{table}[t!]\centering \renewcommand{\arraystretch}{1.5}
	\begin{small}
		\begin{tabular}{||c||c||c||}\hhline{|t:=:t:=:t:=:t|}
					& Gauging String Bargmann Algebra  	& 	Limit of GR: $c \rightarrow \infty$ \\ \hhline{|:=::=::=:|}
			\multirow{2}{*}{Curvature Constraints}
					&	$R_{\mu\nu}^{\phantom{\mu\nu}A} (H) = R_{\mu\nu}^{\phantom{\mu\nu}A'} (P) = 0$
										&	$R_{\mu\nu}^{\phantom{\mu\nu}A} (H) = R_{\mu\nu}^{\phantom{\mu\nu}A'} (P) = 0$ \\ \hhline{||~||-||-||}
					& 	$R_{\mu\nu}^{\phantom{\mu\nu}A} (Z) = 0$
										&	$R_{A'A}{}^A (Z) = R_{A'B'}{}^A (Z) = 0\,$ \\ \hhline{|:=::=::=:|}
			\multirow{2}{*}{Independent Fields}
					&	$\tau_\mu{}^A$ \, $E_\mu{}^{A'}$ \,$m_\mu {}^A$
										&	$\tau_\mu{}^A$ \, $E_\mu{}^{A'}$ \,$m_\mu{}^A$ \\  \hhline{||~||-||-||}
					&	-----				&	$W_{AB}{}^{A'}$ \\  \hhline{|:=::=::=:|}
		 	\multirow{2}{*}{Dependent Fields}
					& 	$\Omega_\mu$ \, $\Omega_\mu{}^{A'B'}$ \,${\Omega}_\mu{}^{AA'}$
										&	$\Omega_\mu$ \, $\Omega_{\mu}{}^{A'B'}$ \,$\tilde{\Omega}_\mu{}^{AA'}$ \\ \hhline{||~||-||-||}
					& 	${n}_\mu$ 		&	-----								 \\ \hhline{|b:=:b:=:b:=:b|}
		\end{tabular}\renewcommand{\arraystretch}{1}
                \caption{This table summarizes the main differences between the geometry obtained by gauging the string Newton-Cartan algebra and the one obtained in the nonrelativistic limit of GR. For the definition of $\tilde \Omega$, see eq.~\eqref{omegatilde}.}
                \label{tb:diffsgauginglimit}
\end{small}
\end{table}
%

The gauging of the string Bargmann algebra thus leads to a version of string Newton-Cartan geometry in which all spin connection fields are fully determined in terms of other fields. This is different from the geometry obtained from the nonrelativistic limit of GR discussed in \S\ref{sec:sncg}. We have summarized the main differences between the results of this Appendix on gauging the string Bargmann algebra and the $c \rightarrow \infty$ limit in \S\ref{sec:sncg} in Table \ref{tb:diffsgauginglimit}.

\section{Gauging the String Newton-Cartan Algebra} \label{app:gaugingext}

In \S\ref{sec:limit}, we found that the string Newton-Cartan geometry derived from the $c \rightarrow \infty$ limit of General Relativity (in the presence of an auxiliary two-form with zero curvature) is less constrained than the one that we derived from gauging the string Bargmann algebra in Appendix \ref{app:gauging}. In particular, there are components $W_{ABA'}$ (defined in \eqref{eq:WABA'}) of the gauge field of string-Galilean boosts that are independent in the limiting procedure but depend on other fields in the gauging procedure. The discussion of \S\ref{sec:gssNC} however indicates that the symmetry algebra that underlies the string Newton-Cartan geometry obtained in the $c\rightarrow \infty$ limit of GR is not the string Bargmann algebra, but rather the string Newton-Cartan algebra. One therefore expects that a gauging of the latter algebra is able to reproduce the main features of string Newton-Cartan geometry as it arises from GR in a nonrelativistic limit. In this Appendix, we show that this expectation is borne out and that the gauging of the string Newton-Cartan algebra indeed leads to a string Newton-Cartan geometry in which $W_{ABA'}$ appear as independent fields.

The gauging of the string Newton-Cartan algebra follows closely that of the string Bargmann algebra given above in Appendix \ref{app:gauging}. The main differences stem from the fact that the commutation relations \eqref{eq:ESNCaGG} are now replaced by the ones of \eqref{eq:commPolextra}. The gauge field $n_\mu$ in \eqref{eq:Thetamu} and the gauge parameter $\sigma$ in \eqref{eq:Pi} are now replaced by a gauge field $n_\mu{}^{AB}$ and a gauge parameter $\sigma^{AB}$, respectively, that obey $n_\mu{}^A{}_A = 0$ and $\sigma^A{}_A = 0$. The gauge transformations of all gauge fields are then given by the rules of \eqref{eq:rules1}, \eqref{eq:conns1} and \eqref{eq:conns2}, as well as the following gauge transformations of $m_\mu{}^A$ and $n_\mu{}^{AB}$
\begin{subequations}
\begin{align}
	\delta m_\mu{}^A & = \p_\mu \sigma^A - \epsilon^A{}_B \sigma^B \Omega_\mu + \Lambda \, \epsilon^A{}_B m_\mu{}^B + \Lambda^{AA'} E_{\mu A'} - \tau_\mu{}^B \sigma^A{}_B\,, \label{eq:modifiedm} \\[4pt]
	\delta n_\mu{}^{AB} & = \p_\mu \sigma^{AB} + 2 \epsilon_C{}^B \bigl( \Omega_\mu \, \sigma^{(AC)} - \Lambda \, n_\mu{}^{(AC)} \bigr) + 2 \Lambda^{[A}{}_{A'} \Omega_\mu{}^{B]A'}\,.
\end{align}
\end{subequations}
that replace the rules derived in \eqref{eq:deltammuA} and \eqref{eq:deltanmu}, respectively.
The curvature two-forms are given by the ones listed in \eqref{eq:curvaturetwoforms}, apart from the ones given \eqref{eq:RmunuAZ} and \eqref{eq:RmunuZ}, that are now replaced by
\begin{subequations}
\begin{align}
	R_{\mu\nu}{}^A (Z) & = 2 \bigl( \p^{}_{[\mu} m^{}_{\nu]}{}^A + \epsilon^A{}_B m^{}_{[\mu}{}^B \Omega^{}_{\nu]} - \tau^{}_{[\mu}{}^B n^{}_{\nu]}{}^A{}_B + E^{}_{[\mu}{}^{A'} \Omega^{}_{\nu]}{}^A{}_{A'}  \bigr)\,, \label{eq:RmunuAZext}\\[4pt]
	R_{\mu\nu}{}^{AB} (Z) & = 2 \bigl( \p^{}_{[\mu} n^{}_{\nu]}{}^{AB} + 2 \epsilon_C{}^B \Omega_{[\mu} n_{\nu]}{}^{(AC)} - \, \Omega^{}_{[\mu}{}^{[A|A'|} \Omega^{}_{\nu]}{}^{B]}{}^{}_{A'} \bigr)\,.
\end{align}
\end{subequations}
Note that $R_{\mu\nu}{}^{AB} (Z)$ also contains a symmetric traceless part,
\be
	R_{\mu\nu}{}^{(AB)} (Z) = 2 \bigl( \p^{}_{[\mu} n^{}_{\nu]}{}^{(AB)} +2 \epsilon_C{}^B \Omega_{[\mu} n_{\nu]}{}^{(AC)} \bigr)\,.
\ee
After imposing the curvature constraints
\begin{align}
	R_{\mu\nu}^{\phantom{\mu\nu}A} (H)
	= R_{\mu\nu}^{\phantom{\mu\nu}A'} (P)
	= R_{\mu\nu}^{\phantom{\mu\nu}A} (Z) = 0\,,
\end{align}
and defining
\begin{align} \label{eq:ellABC}
	\ell_{ABC} \equiv n_{(AB)C} - \frac{1}{2} \eta^{}_{AB} n^D{}_{DC} \,,
		\qquad
	\tilde{n}_\mu{}^{AB} \equiv n_\mu{}^{AB} - \tau_\mu{}^C \ell_C{}^{AB}\,,
\end{align}
we find that one can solve the conventional constraints for all components in $\Omega_\mu$, $\Omega_\mu{}^{A'B'}$, $\Omega_\mu {}^{AA'}$ and $n_\mu{}^{AB}$
in terms of the independent fields $\tau_\mu{}^A$, $E_\mu{}^A$, $m_\mu{}^{A}$, $W_{AB}{}^{A'}$ and $\ell_{ABC}$\,. 
The solutions to the dependent fields $\Omega_\mu$, $\Omega_\mu{}^{A'B'}$, $\tilde{\Omega}_\mu{}^{AA'}$ (defined in \eqref{omegatilde}) and $\tilde{n}_\mu{}^{AB}$ are summarized in Table \ref{tb:counting1} except that the last few rows are now replaced by Table \ref{tb:counting2}.
The explicit expressions for the different components listed in Table \ref{tb:counting2} are given by:
%
%
\begin{table}[t!]
\centering
\begin{tabular}{|| c || c | c ||}\hhline{|t:=:t:==:t|}
\# of components			&${\tilde{\Omega}}_\mu^{\phantom{\mu}AA'}$		&$\tilde{n}_\mu{}^{AB}$						\\\hhline{|:=::==:|}
\textcolor{DB}{$d(d-1)$}
			&\textcolor{DB}{$(d-2)^2$}					&\textcolor{DB}{$3d-4$}						\\\hhline{|:=::==:|}
\multirow{6}{*}{$R_{\mu\nu}^{\phantom{\mu\nu}A} (Z)$}
			&\multirow{2}{*}{\textcolor{DB}{\scalebox{0.9}{0}}}			&$n_{A} {}^{AB}$									\\\hhline{||~||~|-||}
			& 									&\textcolor{DB}{\scalebox{0.9}{2}} 							\\\hhline{||~||-|-||}
			&$\Omega^{A}{}_{AA'}$					&${n}_{A'}{}^{AB}$								\\\hhline{||~||-|-||}
 			&\textcolor{DB}{$d-2$} 					&\textcolor{DB}{$3(d-2)$}				 		\\\hhline{||~||-|-||}
			&$\Omega_{[A'B']A}$					&\multirow{2}{*}{\textcolor{DB}{\scalebox{0.9}{0}}}				\\\cline{2-2}
			&\textcolor{DB}{$(d-2)(d-3)$} 				&			 					  		\\\hhline{|b:=:b:==:b|}
\end{tabular}
\caption{This table shows how to solve  the conventional constraint $R_{\mu\nu}{}^A (Z) = 0$\,. The numbers in the table denote the components of the various ingredients. The decompositions of the conventional constraint equations are listed in eq.~\eqref{eq:solutionscc}.}
\label{tb:counting2}
\end{table}
\begin{subequations} \label{eq:solutionscc}
\begin{align}
	\Omega^A{}_{AA'} & = -2m_{A'A}{}^A\,,
		\qquad
	\Omega_{[A'B']A}  = m_{B'A'A}\,,
		\qquad%
	n^A{}_{AB} = 2 m_{BA}{}^{A}\,, \\[2pt]
	n_{A'AB} & = - W_{ABA'} - 2 \bigl( m_{A'BA} - \tfrac{1}{2} \eta^{}_{AB} m^{}_{A'C}{}^C \bigr) - E_{ABA'}\,. 
\end{align}
\end{subequations}
As desired, now $W_{ABA'}$ appears as an independent gauge field, in terms of which $\Omega_\mu{}^{AA'}$ can be written as 
\begin{align}
	\Omega_\mu{}^{AA'}
	& = - E^{}_\mu{}^{AA'} + E^{}_{\mu B'} E^{AA'\!B'} + m_\mu{}^{A'\!A} + \tau^{}_{\mu B} m^{AA'\!B} \notag \\[2pt]
	& \hspace{1.85cm} + 2 \tau_{\mu B} \bigl[ m^{A'(AB)} - \tfrac{1}{2} \eta^{AB} m^{A'C}{}_C \bigr] + \tau_\mu{}^B W_{B}{}^{AA'}\,,
\end{align}
which agrees with \eqref{eq:OmegamuAA'}.

Note that this gauging procedure of the string Newton-Cartan algebra does not only lead to the same transformation rules for the independent fields $\tau_\mu{}^A$, $E_\mu{}^{A^\prime}$ and $m_\mu{}^A$, that appeared in the limit of GR discussed in this paper. The gauging also reproduces the expressions for the spin connections $\Omega_\mu$, $\Omega_\mu{}^{A^\prime B^\prime}$ and $\Omega_\mu{}^{A A^\prime}$ that were found in this limit. In particular, the same components $W_{AB A^\prime}$ of the string-Galilean boost connection are left undetermined, while all other components and spin connections are determined in terms of other fields by the same expressions that follow from the limit. In this sense, the gauging of the string Newton-Cartan algebra can be viewed as an alternative way to construct the string Newton-Cartan geometry coming from GR. Note however that, compared to the limit, the gauging gives rise to an extra gauge field $n_\mu{}^{AB}$, that is only partially dependent and that is absent in both the limits of the kinematics of GR and of the relativistic string action.

In \eqref{eq:Dalg}, we introduced an extension of the string Newton-Cartan algebra with a dilatational generator $D$\,. In the gauging procedure, one assigns an extra gauge field $\Phi_\mu$ to this dilatational generator. First, we list some of the important modifications to the transformation rules. Under the dilatational transformation parametrized by $\Lambda_D$\,, we have
\be
    \delta_{D} \tau_\mu{}^A = \Lambda_D \, \tau_\mu{}^A\,, 
        \qquad
    \delta_D \Phi_\mu = \p_\mu \Lambda_D\,,
        \qquad
    \delta_D m_\mu{}^A = - \Lambda_D \, m_\mu{}^A\,.
\ee
Under the $Z_A$ transformation, we have
\be \label{eq:dmPhiunprimed}
    \delta m_\mu{}^A = \p_\mu \sigma^A - \epsilon^A{}_B \, \sigma^B \Omega_\mu + \Phi_\mu \sigma^A\,.
\ee
Second, the curvature two-tensors are also made dilatational covariant. For example, the curvature constraint associated with $H_A$ now reads
\be \label{eq:RHPhiunprimed}
    R_{\mu\nu}{}^A (H) = 2 \bigl( \p_{[\mu} + \Phi_{[\mu} \bigr) \tau_{\nu]}{}^A + \epsilon^A{}_B \, \tau_{[\mu}{}^B \Omega_{\nu]} \,.
\ee

One can realize this extended algebra in the worldsheet action in terms of different gauge fields $\tau'_\mu{}^A$\,, $E'_\mu{}^{A'}$ and $m'_\mu{}^A$ than the ones used in \eqref{eq:nraction}
with
\begin{align}
    \tau'_\mu{}^A = e^{-\Phi} \, \tau_\mu{}^A\,, 
        \qquad %
    E'_\mu{}^{A'} = E_\mu{}^{A'}\,,
        \qquad %
    m'_\mu{}^A = e^{\Phi} \, m_\mu{}^A\,,
\end{align}
with $\tau_\mu{}^A$\,, $E_\mu{}^{A'}$ and $m_\mu{}^A$ the field variables used throughout the bulk of the paper.
Then, the string action \eqref{eq:nraction} becomes
\begin{align}
        S_G & = \frac{1}{4\pi\alpha'} \int d^2 \sigma \sqrt{h} \, \Bigl[ \mathcal{D} x^\mu \overline{\mathcal{D}} x^\nu \bigl( H'_{\mu\nu} + B_{\mu\nu} \bigr) + \lambda \, \overline{\mathcal{D}} x^\mu \tau'_\mu \, e^\Phi + \overline{\lambda} \, {\mathcal{D}} x^\mu \overline{\tau}'_\mu \, e^\Phi \Bigr] \notag \\
		& \quad + \frac{1}{4\pi} \int d^2 \sigma \sqrt{h} \, R^{(2)} \, \bigl(- \tfrac{1}{4} \ln G' \bigr)\,,
\end{align}
where
\begin{subequations}
\begin{align}
    H'_{\mu\nu} & = E'_\mu{}^{A'} E'_\nu{}^{A'} + \bigl( \tau'_\mu{}^A m'_\nu{}^B + \tau'_\nu{}^A m'_\mu{}^B \bigr) \eta_{AB}\,, \\[2pt]
    G' & = \det^{(d)} \bigl( {H'}_{\mu\nu} \bigr) \det^{(2)} \bigl( \tau'_\rho{}^A {H'}^{\rho\sigma} \tau'_\sigma{}^B \bigr)\,.
\end{align}
\end{subequations}
In this basis, the $Z_A$ symmetry acts on $m'_\mu{}^A$ as
\be \label{eq:ZAPhim}
    \delta_{Z_A} m'_\mu{}^A = \bigl( \p_\mu + \p_\mu \Phi \bigr) \sigma^A - \epsilon^A{}_B \, \Omega'_\mu{} \sigma^B\,,
\ee
provided the following transformation rule is assigned to $\lambda$ and $\overline{\lambda}$\,:
\be
    \delta_{Z_A} \lambda = e^{-\Phi} \, \mathcal{D} x^\mu \bigl( \p_\mu - \Omega'_\mu + \p_\mu \Phi \bigr) \overline{\sigma}\,,
        \quad
    \delta_{Z_A} \overline{\lambda} = e^{-\Phi} \, \overline{\mathcal{D}} x^\mu \bigl( \p_\mu + \Omega'_\mu + \p_\mu \Phi \bigr) \sigma\,,
\ee
and provided the curvature constraint
\be \label{eq:ccPhi}
     \p^{}_{[\mu} {\tau'}^{}_{\nu]}{}^{A} + \epsilon^A{}_{B} {\tau'}^{}_{[\mu}{}^B {\Omega'}^{}_{\nu]} = - \tau'_{[\mu} \p_{\nu]} \Phi\,.
\ee
Note that \eqref{eq:ZAPhim} and \eqref{eq:ccPhi} are natural from the gauging perspective, if one views the primed fields as gauge fields of the string Newton-Cartan algebra extended by a dilatational generator and applies the constraint 
\be
R_{\mu\nu}(D) \equiv 2 \partial_{[\mu} \Phi_{\nu]} =0 \,,
\ee
which can then be locally solved by $\Phi_\mu = \partial_\mu \Phi$. In that case, \eqref{eq:ZAPhim} coincides with \eqref{eq:dmPhiunprimed} and \eqref{eq:ccPhi} coincides with \eqref{eq:RHPhiunprimed}.

\newpage

\bibliographystyle{JHEP}
\bibliography{geonrs}

\end{document}